\documentclass[11pt]{article}

\usepackage[T1]{fontenc}
\usepackage[utf8]{inputenc}
\usepackage{epsfig}
\usepackage{amsmath,amssymb,amsthm,amsfonts}
\usepackage{a4}
\usepackage{tikz}
\usetikzlibrary{positioning}
\tikzset{main node/.style={circle,fill=blue!20,draw,minimum size=1cm,inner sep=0pt}}
\usetikzlibrary{arrows.meta}
\usepackage{color}
\usepackage{natbib}
\usepackage{graphicx}       
\usepackage{ccaption}
\usepackage[labelformat=simple]{subcaption}

\usepackage{listings}
\usepackage{listings}
\usepackage{enumerate}
\usepackage{wrapfig}
\usepackage{pgfplots}

\usepackage{ccaption}
\captionnamefont{\bfseries}

\bibpunct{(}{)}{;}{a}{}{,}

\begin{document}
\begin{center}
{\noindent{\LARGE \textbf{A multiple-try Metropolis--Hastings algorithm with tailored proposals} \vspace{1cm} \\} 
{\Large \textsc{Xin Luo}}\\{\it Department of
    Mathematical Sciences, Norwegian University of Science and
    Technology}\vspace{1cm}\\{\Large \textsc{H\aa kon Tjelmeland}}\\{\it Department of
    Mathematical Sciences, Norwegian University of Science and
    Technology}\vspace{1cm}
}	
\end{center}

\begin{abstract} 
We present a new multiple-try Metropolis--Hastings algorithm designed to be especially beneficial when 
a tailored proposal distribution is available. The algorithm is based on a given acyclic graph $\mathcal{G}$,
where one of the nodes in $\mathcal{G}$, $k$ say, contains the current state of the Markov chain and the remaining nodes
contain proposed states generated by applying the tailored proposal distribution. The Metropolis--Hastings
algorithm alternates between two types of updates. The first update type is using the tailored proposal 
distribution to generate new states in all nodes in $\mathcal{G}$ except in node $k$. The second update
type is generating a new value for $k$, thereby changing the value of the current state. We evaluate
the effectiveness of the proposed scheme in an example with previously defined target and proposal distributions.
\end{abstract}
 
\vspace{0.5cm}

\vspace{-0.1cm}

\section{Introduction}
In the field of Bayesian inference, a popular and powerful tool is Markov chain Monte Carlo (MCMC) methods 
\citep{book17, book24, book30}. This includes the Gibbs sampler \citep{art13} and the Metropolis-Hastings (MH) algorithm 
\citep{art71, art11}, where the former is a special case of the latter. In principle, any 
distribution $p(x)$ that can be evaluated up to a normalizing constant, can be simulated using the Metropolis--Hastings setup. 
The algorithm is iterative with each iteration consisting of two parts. 
Letting $x$ denote the current state, first a potential new state $\widetilde{x}$ is generated from a proposal 
distribution $q(\widetilde{x}|x)$, and thereafter $\widetilde{x}$
is accepted with probability
\begin{equation}
\alpha (x|\widetilde{x}) = \min\left\{ 1,\frac{p(\widetilde{x})q(x|\widetilde{x})}{p(x)q(\widetilde{x}|x)}\right\}
\end{equation}
and otherwise the current state $x$ is retained. The choice of the proposal distribution $q(\widetilde{x}|x)$ is essential for the 
convergence and mixing properties of the simulated Markov chain, and therefore for the computation time necessary for 
exploring the target distribution $p(x)$. Often very simple proposal distributions are adopted, with a Gaussian centered at 
the current state $x$ and full conditional distributions being the most popular ones, and for many target distributions
this is sufficient to get acceptable convergence and mixing properties. For other target distributions $p(x)$, however, 
such choices give too slow convergence and mixing for the algorithm to be practical. In the literature different
remedies have been proposed to cope with such a situation. In principle, a simple solution is to tailor
the proposal distribution to the specific target distribution in question. The Metropolis--Hastings setup is very general,
and in particular the proposal distribution is allowed to depend on properties of the target distribution. Thereby one may 
let $q(\widetilde{x}|x)$ depend on properties of the target distributions $p(x)$ close to, in some sense, the current state $x$. 
Such a tailored proposal distribution may dramatically reduce the number of iterations to get convergence and sufficient 
mixing, and may therefore be beneficial even if simulating from such a tailored proposal distribution typically requires a 
lot more computation time than sampling from one of the simple proposal distributions discussed above. Examples of such 
tailored proposal distributions can for example be found in \citet{art43}, \citet{art165} and \citet{tech34}.

\citet{art74} introduces an alternative strategy for coping with a target distribution where the use of token proposal distributions
do not give sufficiently good convergence and mixing. An generalized version of the Metropolis--Hastings algorithm is proposed, 
called the Multiple-try Metropolis (MTM) algorithm. Also this algorithm is based on a proposal distribution $q(\widetilde{x}|x)$,
but instead of proposing only one potential new state in each iteration, MTM generates several potential 
new states from the proposal distribution. The potential new states are  generated conditionally independent given the 
current state $x$. Next, one of the potential new states is, with a certain probability for each potential state, chosen as the 
next current state, or all the proposed states are rejected and the current state thereby retained.
The idea is that by generating several potential states one can better explore the sample space and thereby better
convergence and mixing can be obtained. Moreover, as the potential new states are generated independently given the current state,
the generation of the potential new states can be parallelized. Intuitively one should expect the performance of 
MTM to improve as the number of potential new states increases. \citet{art159} show, however, 
that there are cases where the performance does not improve when increasing the number of potential new states. 
Many variants of the MTM algorithm have later been proposed, see in particular \citet{art79}, \citet{art160}, 
\citet{pro23}, \citet{art162} and \citet{art163}. The use of MTM is also discussed in \citet{art164}, \citet{art161} and \citet{tech35}.

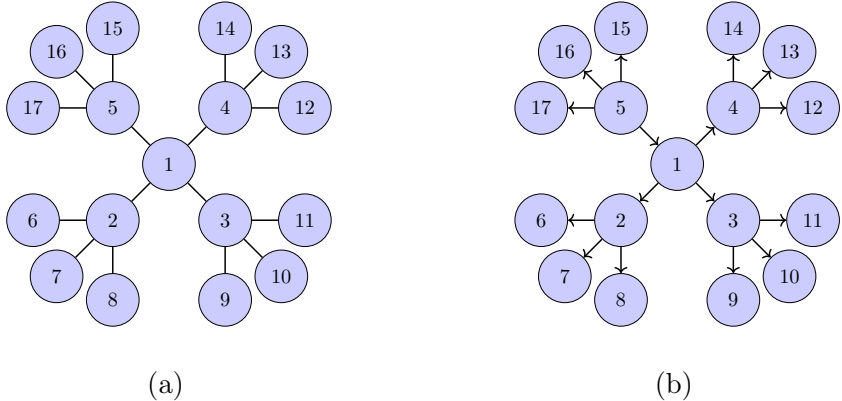
\begin{figure}
\begin{center}
\begin{tabular}{ccc}
\scalebox{0.7}{
\begin{tikzpicture}
\node[main node] (1) {$1$};
\node[main node] (2) [below left = 0.35cm and 0.35cm of 1] {$2$};
\node[main node] (3) [below right = 0.35cm and 0.35cm of 1] {$3$};
\node[main node] (4) [above right = 0.35cm and 0.35cm of 1] {$4$};
\node[main node] (5) [above left = 0.35cm and 0.35cm of 1] {$5$};

\node[main node] (6) [left = 0.5cm of 2] {$6$};
\node[main node] (7) [below left = 0.35cm and 0.35cm of 2] {$7$};
\node[main node] (8) [below = 0.5cm of 2] {$8$};
\node[main node] (9) [below = 0.5cm of 3] {$9$};
\node[main node] (10) [below right = 0.35cm and 0.35cm of 3] {$10$};
\node[main node] (11) [right = 0.5cm of 3] {$11$};
\node[main node] (12) [right = 0.5cm of 4] {$12$};
\node[main node] (13) [above right = 0.35cm and 0.35cm of 4] {$13$};
\node[main node] (14) [above = 0.5cm of 4] {$14$};
\node[main node] (15) [above = 0.5cm of 5] {$15$};
\node[main node] (16) [above left = 0.35cm and 0.35cm of 5] {$16$};
\node[main node] (17) [left = 0.5cm of 5] {$17$};

\path[draw,thick]
(1) edge node {} (2)
(1) edge node {} (3)
(1) edge node {} (4)
(1) edge node {} (5)
(2) edge node {} (6)
(2) edge node {} (7)
(2) edge node {} (8)
(3) edge node {} (9)
(3) edge node {} (10)
(3) edge node {} (11)
(4) edge node {} (12)
(4) edge node {} (13)
(4) edge node {} (14)
(5) edge node {} (15)
(5) edge node {} (16)
(5) edge node {} (17);

\end{tikzpicture}}
& \hspace*{1.5cm} &
\scalebox{0.7}{
\begin{tikzpicture}
\node[main node] (1) {$1$};
\node[main node] (2) [below left = 0.35cm and 0.35cm of 1] {$2$};
\node[main node] (3) [below right = 0.35cm and 0.35cm of 1] {$3$};
\node[main node] (4) [above right = 0.35cm and 0.35cm of 1] {$4$};
\node[main node] (5) [above left = 0.35cm and 0.35cm of 1] {$5$};

\node[main node] (6) [left = 0.5cm of 2] {$6$};
\node[main node] (7) [below left = 0.35cm and 0.35cm of 2] {$7$};
\node[main node] (8) [below = 0.5cm of 2] {$8$};
\node[main node] (9) [below = 0.5cm of 3] {$9$};
\node[main node] (10) [below right = 0.35cm and 0.35cm of 3] {$10$};
\node[main node] (11) [right = 0.5cm of 3] {$11$};
\node[main node] (12) [right = 0.5cm of 4] {$12$};
\node[main node] (13) [above right = 0.35cm and 0.35cm of 4] {$13$};
\node[main node] (14) [above = 0.5cm of 4] {$14$};
\node[main node] (15) [above = 0.5cm of 5] {$15$};
\node[main node] (16) [above left = 0.35cm and 0.35cm of 5] {$16$};
\node[main node] (17) [left = 0.5cm of 5] {$17$};

\path[->,thick]
(1) edge node {} (2)
(1) edge node {} (3)
(1) edge node {} (4)
(5) edge node {} (1)
(2) edge node {} (6)
(2) edge node {} (7)
(2) edge node {} (8)
(3) edge node {} (9)
(3) edge node {} (10)
(3) edge node {} (11)
(4) edge node {} (12)
(4) edge node {} (13)
(4) edge node {} (14)
(5) edge node {} (15)
(5) edge node {} (16)
(5) edge node {} (17);

\end{tikzpicture}}\\[0.4cm]
(a) & & (b)
\end{tabular}
\end{center}
\caption{\label{fig:graph}(a) An acyclic graph with $n=17$ nodes, where the nodes are numbered from $1$ to $17$. (b) The 
directed acyclic graph resulting from the graph in (a) when $k=5$.}
\end{figure}

In this article we propose a setup which combines the two approaches discussed above. The starting point of our
scheme is an acyclic graph with $n$ nodes, where the nodes are numbered from $1$ to $n$. 
A small example graph is shown in Figure \ref{fig:graph}(a). 
To each node $i$ in the graph we associate a random quantity $x_i$ with the same sample space as the target distribution 
$f(x)$. We also introduce a random index $k\in\{1,\ldots,n\}$ with a distribution $f(k)$, which may be the 
uniform distribution. Given a value for $k$ we assume the distribution of $x_k$ to equal the target distribution.
Next, given $x_k$ the distribution of the remaining variables $x_i,i\neq k$ is defined by a proposal distribution
$q(\cdot|\cdot)$. For example, if the acyclic graph is the one in Figure \ref{fig:graph}(a) and $k=5$,
we assume $x_1$, $x_{15}$, $x_{16}$ and $x_{17}$ be be conditionally independent and identically distributed according to 
$q(\cdot |x_5)$, and in the next step we assume $x_2$, $x_3$ and $x_4$ to be conditionally independent and independently 
distributed according to $q(\cdot |x_1)$, and so on. We end up with the directed acyclic graph (DAG)
shown in Figure \ref{fig:graph}(b), where all the directed edges represent the same proposal distribution.
We have thereby defined a joint distribution for $k$ and $x_1,\ldots,x_n$ and can adopt a 
Metropolis--Hastings algorithm to simulate from this joint distribution. By construction the conditional 
distribution of $x_k$ given $k$ is equal to the target distribution so by simulating from the joint 
distribution we also get samples from the target distribution of interest. Even if we define a joint 
distribution for all $k,x_1,\ldots,x_n$, the $x_i,i\neq k$ is best considered as proposed potential 
new states. One should note that the potential new states are not conditionally independent given $x_k$ 
as are common in MTM methods, so in this sense our setup defines a generalized MTM scheme. Even if our
setup is well defined for any proposal distribution $q(\cdot|\cdot)$, we should only expect 
favorable results with the procedure when $q(\cdot|\cdot)$ is tailored to the specific target distribution 
of interest. If, for example, a simple random walk proposal is adopted and the graph is as shown in 
Figure \ref{fig:graph} with $k=5$, one should expect the higher order proposals in
$x_{6}, x_{7}, \ldots,x_{14}$ to be in the tail of the target distribution and thereby to have low acceptance probabilities.
With a proposal distribution tailored to the specific target distribution, however, also higher order
proposals should have a reasonable chance of getting high acceptance probabilities.

The remainder of this article is organized as follows. In section \ref{sec:algorithm} we specify and present the 
mathematical details for our proposed multiple-try Metropolis--Hastings algorithm, assuming the sample space of 
the target distribution to be of a fixed dimension. In Section \ref{sec:reversiblejump} we generalize the setup \
to a situation where the sample space is allowed to be of varying dimension, so that a reversible jump
proposal must be used. In Section \ref{sec:examples} we present the results of some simulation examples, 
and finally we give some closing remarks in Section \ref{sec:closingremarks}.

\section{\label{sec:algorithm}The algorithm}
As in the setup of a standard MH algorithm, we let $p(x), x\in\mathbb{R}^m$ denote the target distribution and $q(\widetilde{x}|x)$ a proposal distribution from state $x$ to state $\widetilde{x}$, where $x,\widetilde{x}\in\mathbb{R}^m$. In addition, the algorithm is based on a chosen connected undirected acyclic labeled graph $\mathcal{G}=(\mathcal{V}, \mathcal{E})$ with $n > 1$ vertices, where $\mathcal{V}=\{1,\dotsc,n\}$ is the set of vertices and $\mathcal{E}\subset\{\{i,j\}|i,j\in\mathcal{V},i\neq j\}$ is the set of undirected edges. An example of such a graph with $n=17$  vertices is illustrated in Figure \ref{fig:graph}(a). Note that the notation $\{i,j\}$ with a pair of curly braces represents an undirected edge connecting vertex $i$ and vertex $j$. Given the graph $\mathcal{G}$, for each $k\in\mathcal{V}$ let $\mathcal{G}_k=(\mathcal{V}, \mathcal{E}_k)$ be the DAG obtained from $\mathcal{G}$ by defining vertex $k$ to be a root vertex and letting all edges be oriented away from this root. Thus, $\mathcal{E}_k\subset\{(i,j)|i,j\in\mathcal{V},i\neq j\}$, where the notation $(i,j)$ with a pair of parentheses represents an edge in the direction from vertex $i$ to vertex $j$. Figure \ref{fig:graph}(b) depicts the resulting $\mathcal{G}_5$ when $\mathcal{G}$ is as shown in Figure \ref{fig:graph}(a). 

To each $i\in\mathcal{V}$ we associate a stochastic variable $x_i\in\mathbb{R}^m$. We also define a discrete stochastic variable $k\in \mathcal{V}$, which we assume to be uniformly distributed. Given $k$ we assume $x_k$ to be distributed according to the target distribution, i.e. $f(x_k|k)=p(x_k)$. Next, given $x_k$ the distribution of the remaining variables $x_i,i\neq k$ is defined by the graph $\mathcal{G}_k$ and the proposal distribution $q(\cdot|\cdot)$. More specifically, we assume the $x_i$'s to have a Markov property as specified by $\mathcal{G}_k$, and for each $(i,j)\in\mathcal{E}_{k}$ we assume $x_j|x_i\sim q(x_j|x_i)$. Thereby the joint distribution of $k$ and $x_i,i\in \mathcal{V}$ becomes
\begin{equation}\label{eq:joint}
f(k,x_1,\dotsc,x_n)=\frac{1}{n}\cdot p(x_k)\prod_{(i,j)\in\mathcal{E}_k}q(x_j|x_i).
\end{equation}
Note that by construction $f(k,x_k)=f(k)p(x_k)$. Thereby we have $f(x_k|k)\propto f(k,x_k)\propto p(x_k)$. This means that we can obtain a sample from the target distribution $p(x)$ by first producing a sample $(k,x_1,\dotsc,x_n)$ from \eqref{eq:joint} and thereafter picking out $x_k$.

We now discuss how to simulate from the distribution given in \eqref{eq:joint}. We choose to draw the values of $k$ and $\{x_i|i\in\mathcal{V},i\neq k\}$ in turn by Gibbs updates. The full conditional for $\{x_i|i\in\mathcal{V},i\neq k\}$ is clearly
\begin{equation}\label{eq:full cond1}
f(x_1,\dotsc,x_{k-1},x_{k+1},\dotsc,x_n|k,x_k)=\prod_{(i,j)\in\mathcal{E}_k}q(x_j|x_i).
\end{equation}
We simulate the new values for $\{x_i|i\in\mathcal{V},i\neq k\}$ in the order specified by $\mathcal{G}_k$. For the graph in Figure 1(b) for examples, $k=5$ so we first sample $x_1$, $x_{15}$, $x_{16}$ and $x_{17}$ given $x_5$, thereafter we can sample $x_2$, $x_3$ and $x_4$ given $x_1$, and finally we can sample $x_6$, $x_7$ and $x_8$ given $x_2$, $x_9$, $x_{10}$ and $x_{11}$ given $x_3$, and $x_{12}$, $x_{13}$ and $x_{14}$ given $x_4$.

The full conditional distribution for $k$ becomes 
\begin{equation}\label{eq:full cond2}
f(k|x_1,\dotsc,x_n)=\dfrac{p(x_k)\prod_{(i,j)\in\mathcal{E}_k}q(x_j|x_i)}{\sum_{r=1}^{n}\left[p(x_r)\prod_{(i,j)\in\mathcal{E}_r}q(x_j|x_i)\right]}.
\end{equation}
Since $k$ is a discrete variable, we readily sample the new value of $k$ by first computing the probability for each possible value of $k$, and thereafter applying the standard algorithm for sampling from a discrete distribution, see for example \citet{book30}.

Note that in the above setup we specify the simulation algorithm by choosing the proposal distribution $q(\cdot|\cdot)$ and the graph $\mathcal{G}$. By choosing a graph with many vertices we get an algorithm where a large number of potential new states are proposed in each iteration, and by choosing a graph with long paths some of the proposed states may differ a lot from the current state $x_k$. If we use the graph in Figure \ref{fig:graph} and $k=5$ for example, the potential new states $x_6$ to $x_{14}$ are generated by applying the proposal distribution $q(\cdot|\cdot)$ three times. 
As also discussed in the introduction, it is not reasonable to combine such a graph with a simple random walk proposal $q(\cdot|\cdot)$, since applying such a $q(\cdot|\cdot)$ several times will just leave us 
in some tail of the target distribution $p(\cdot)$. Adopting a more tailored proposal mechanism, however, 
we can obtain high probability proposals even after having iterated the proposal mechanism several times. Clearly the computation time necessary for each iteration of the procedure proposed above depends on the number of vertices in the graph, so choosing a good graph $\mathcal{G}$ is a trade-off between the possibility for large changes in the state vector in each iteration and required computation time for each iteration. We expect that the better tailored the proposal distribution is to the target distribution, the larger the graph and the longer the paths of the graph should be.

Furthermore, the algorithm can be implemented in parallel not only in sequence.  Based on the structure of the graph, it gives the property of conditional independence, so given a vertex the vertices conditioned on it can be sampled in parallel. For example in Figure \ref{fig:dag}(b), given vertex $5$ we can sample vertices $1$, $15$, $16$ and $17$ in parallel. Given vertex $1$ we can then sample vertices $2$, $3$ and $4$ in parallel, and so on until all vertices are sampled.

In the above we have assumed the dimension, $m$, of the state vector to fixed. In the next section we generalize the setup to a situation where the sample space of the state vector is a union of spaces of different dimensions, i.e. to the reversible jump \citep{art22} situation. For each edge $(i,j)\in \mathcal{G}_k$ a new state is then proposed as in the reversible jump setup. The basic simulation procedure remains the same, but the mathematical details become different.

\section{\label{sec:reversiblejump}The algorithm with a reversible jump proposal distribution}
Let $p(x); x\in \mathcal{X}$ denote the target distribution of interest, where the sample space $\mathcal{X}$ may be a union of spaces
of different dimensions. As in the standard reversible jump setup \citep{art22}. More specifically, we first generate a potential new state
$\widetilde{x}$ by first proposing a variable $u\in\mathcal{U}$ from a proposal distribution $q(u|x);x\in\mathcal{X},u\in\mathcal{U}$, 
where the sample space $\mathcal{U}$ also may be a union of spaces of different dimensions. Next, the potential new state $\widetilde{x}$
is given by some deterministic function of $x$ and $u$, $\widetilde{x}=h(x,u)$ say. Moreover, we have a deterministic function
of $x$ and $u$ which returns a $\widetilde{u}\in\mathcal{U}$, $\widetilde{u}=g(x,u)$ say, so that we have the one-to-one
relation 
\begin{equation}\label{eq:onetoone}
  \left. 
    \begin{array}{r}
      \widetilde{x}=g(x,u) \\
      \widetilde{u}=h(x,u) 
    \end{array}\right\}
\mbox{  }  \Leftrightarrow \mbox{  }
  \left\{
    \begin{array}{r}
      x=g(\widetilde{x},\widetilde{u}) \\
      u=h(\widetilde{x},\widetilde{u}) 
    \end{array}
  \right.
\end{equation}
for any $x,\widetilde{x}\in\mathcal{X}$ and $u,\widetilde{u}\in\mathcal{U}$. As usual in the reversible jump setting the dimension 
matching criterion must be met, i.e. $\dim(x)+\dim(u)=\dim(\widetilde{x})+\dim(\widetilde{u})$. The Jacobian determinant of the 
transformation from $(x,u)$ to $(\widetilde{x},\widetilde{u})$ we denote by 
\begin{equation}\label{eq:jacobian}
J(x,u) = \left|
\begin{array}{cc} \dfrac{\partial g}{\partial x}(x,u) & \dfrac{\partial g}{\partial u}(x,u) \\[0.5cm]
\dfrac{\partial h}{\partial x}(x,u) & \dfrac{\partial h}{\partial u}(x,u)\\[0.1cm] \end{array}\right|.
\end{equation}
Note that the one-to-one relation in (\ref{eq:onetoone}) implies that 
\begin{equation}\label{eq:jacobianInverse}
J(\widetilde{x},\widetilde{u})=J(x,u)^{-1}.
\end{equation}

In addition to the target distribution $p(x)$, the proposal distribution $q(u|x)$ and the one-to-one relation in (\ref{eq:onetoone}), 
our algorithm here is as in the previous section based on a chosen undirected acyclic labeled 
graph $\mathcal{G}=(\mathcal{V},\mathcal{E})$ with $n>1$ vertices, where $\mathcal{V}=\{1,\ldots,n\}$ and $\mathcal{E}$ is the 
set of undirected edges. As in the previous section we also let $\mathcal{G}_k = (\mathcal{V},\mathcal{E}_k)$ denote 
the DAG resulting from $G$ by defining the vertex $k\in\mathcal{V}$ to be a root.
To each $i\in \mathcal{V}$ we again associate a stochastic variable $x_i\in\mathcal{X}$, define a uniformly distributed 
discrete stochastic variable $k\in\mathcal{V}$, and given $k$ we assume $x_k\sim p(x_k)$ so that $f(x_k|k)=p(x_k)$.
Given $k$ and $x_k$ the distribution for the remaining $x_i,i\neq k$ is defined by the DAG $\mathcal{G}_k$ and the 
reversible jump proposal mechanism discussed above. Thus, to each directed edge $(i,j)\in\mathcal{E}_k$ we have a 
$u_{(i,j)}\in\mathcal{U}$ where $u_{(i,j)}|x_i \sim q(u_{(i,j)}|x_i)$ and $x_j=g(x_i,u_{(i,j)})$. One should note that to each
directed edge $(i,j)\in \mathcal{E}_k$ we also have a $u_{(j,i)}=h(x_j,u_{(i,j)})$ which can be used to take us from $x_j$
to $x_i=g(x_i,u_{(j,i)})$ if the direction of the edge is reversed.

Before defining an MCMC algorithm able to simulate the variables discussed above we need to formulate the joint 
distribution for the variables involved. This is, however, difficult when using the above notation. The distribution is 
specified by $k\sim \mbox{Uniform}(\mathcal{V})$, $x_k|k\sim p(x_k)$ and $u_{(i,j)}|x_i\sim q(u_{(i,j)}|x_i)$, but 
formulated in this way the value of $k$ decides not only the distribution of the remaining variables, but also 
what variables that are involved in the specification. We therefore need a new notation where only the values
of the variables involved change with the value of $k$. Therefore, let $x\in \mathcal{X}$, without a subscript, be equal to $x_k$,
whatever values $k$ have, and let $u_{\{i,j\}}\in\mathcal{U}$ be equal to $u_{(i,j)}$ if $(i,j)\in\mathcal{G}_k$ and 
equal to $u_{(j,i)}$ if $(j,i)\in\mathcal{G}_k$. The joint distribution of interest can then be formulated as
\begin{equation}\label{eq:joint_rj}
f(k,x,\{u_{\{i,j\}}|\{i,j\}\in\mathcal{E}\})=\frac{1}{n}\cdot p(x)\prod_{(i,j)\in\mathcal{E}_k}q(u_{\{i,j\}}|x_i).
\end{equation}

To simulate from (\ref{eq:joint_rj}) we basically adopt the same strategy as we did in the previous section, we update 
$k$ and $\{u_{\{i,j\}},\{i,j\}\in\mathcal{E}\}$ in turn by Gibbs updates. For updating $\{u_{\{i,j\}}|\{i,j\}\in\mathcal{E}\}$, the full 
conditional is simply
\begin{equation}
f(\{u_{\{i,j\}}|\{i,j\}\in\mathcal{E}\}|k,x)\propto\prod_{(i,j)\in\mathcal{E}_k}q(u_{\{i,j\}}|x_i).
\end{equation}
This implies that we can sample the new values of the $u_{\{i,j\}}$'s sequentially according to $\mathcal{G}_k$. 
For example, regarding the case in 
Figure \ref{fig:graph}(b) where $k=5$, we first independently sample $u_{(5,1)}, u_{(5,15)}, , u_{(5,16)}$ and $u_{(5,17)}$ given $x_5$ from $q(\cdot|x_5)$, 
and compute $x_1=g(x_5,u_{(5,1)}), x_{15}=g(x_5,u_{(5,15)}), , x_{16}=g(x_5,u_{(5,16)})$ and $x_{17}=g(x_5,u_{(5,17)})$. Thereafter we independently sample 
$u_{(1,2)}, u_{(1,3)}$ and $u_{(1,4)}$ given $x_1$ from $q(\cdot|x_1)$, and compute
$x_2=g(x_1,u_{(1,2)}), x_3=g(x_1,u_{(1,3)})$ and $x_4=g(x_1,u_{(1,4)})$, and so on until we have sampled new values
for all $u_{\{i,j\}}$ and obtained new values for all $x_i,i\neq k$.

When updating $k$ we keep all $x_i,i\in \mathcal{V}$ and $u_{(i,j)},u_{(j,i)}$ for $\{i,j\}\in\mathcal{V}$ fixed. One should note
this implies that the variables $x$ and $u_{\{i,j\}},\{i,j\}\in\mathcal{E}$ used to formulate the joint distribution in 
(\ref{eq:joint_rj}) may change. Arbitrarily choosing $k=1$ as a base case, we choose to sample the new value of $k$ independent of its
current value from the distribution
\begin{equation}\label{eq:prop_k}
r(k)=\dfrac{p(x_{k})\prod_{(i,j)\in\mathcal{E}_{k}}q(u_{(i,j)}|x_i)\prod_{(i,j)\in\mathcal{E}_1\setminus\mathcal{E}_{k}}|J(x_i,u_{(i,j)})|}{\sum_{l\in\mathcal{V}}\left[p(x_l)\prod_{(i,j)\in\mathcal{E}_l}q(u_{(i,j)}|x_i)\prod_{(i,j)\in\mathcal{E}_1\setminus\mathcal{E}_l}|J(x_i,u_{(i,j)})|\right]}, \mbox{ }k\in\mathcal{V}.
\end{equation}
One should note that choosing another base case, for example substitute $\mathcal{E}_1$ with $\mathcal{E}_3$ 
in the above formula, will not not change the distribution $r(k)$. The effect of such a change 
is just to multiply the numerator and the denominator with the same product of Jacobians. 
Since $k$ is a discrete variable it is easy to sample from $r(k)$ by, just in Section \ref{sec:algorithm}, 
applying the standard algorithm for sampling from a discrete distribution.

We show in Appendix \ref{app:acceptance} that the Metropolis--Hastings reversible jump acceptance probability
when using proposal distribution $r(k)$ is identical to one. To update $k$ by sampling the new
value from (\ref{eq:prop_k}) can therefore best to thought of as a Gibbs update. Just as in  a standard
Gibbs update the proposed value is generated independently of the current value and the proposed 
value is always accepted.

\section{\label{sec:examples}Simulation examples}
In the geostatistical community it has over the past years become common practice to estimate a prior model for the 
spatial distribution of reservoir properties from one or several training images. A training image 
can be an observed or constructed scene of a discrete variable defined in a rectangular lattice, see for
example \citet{book41} and references therein. \citet{tech34} introduce a Markov mesh model (MMM) and a corresponding tailored 
proposal distribution for such a situation. Here the target distribution $p(x)$ is the posterior of the 
model parameters in a Markov mesh model \citep{art133,art119} when conditioning on a training image. In the following we first 
discuss the target distribution $p(x)$ and the corresponding tailored proposal distribution, and thereafter
present simulation results for two different training images.

\subsection{\label{subsec:target}The target distribution}
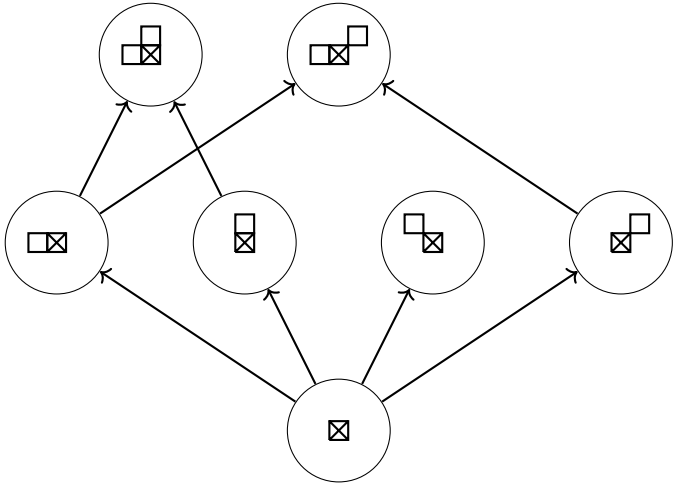
\begin{figure}
\begin{center}
\begin{tikzpicture}[scale=2.5]
  \def\radius{38.9};
  \coordinate (Pempty) at (0,0);
  \coordinate (P1) at (-1.5,1);
  \coordinate (P2) at (-0.5,1);
  \coordinate (P3) at (0.5,1);
  \coordinate (P4) at (1.5,1);
  \coordinate (P12) at (-1,2);
  \coordinate (P14) at (0,2);

  \coordinate (c1) at (-0.05,-0.05);
  \coordinate (c2) at (0.05,-0.05);
  \coordinate (c3) at (0.05,0.05);
  \coordinate (c4) at (-0.05,0.05);

  \coordinate (r1) at (-0.1,0);
  \coordinate (r2) at (0,0.1);
  \coordinate (r3) at (-0.1,0.1);
  \coordinate (r4) at (0.1,0.1);

  \node[draw,circle,inner sep=0pt,minimum size=\radius,name=Nempty] at (Pempty) {};
  \draw[thick] (Pempty) +(c1) -- +(c2) -- +(c3) -- +(c4) -- +(c1) +(c1) -- +(c3) +(c2) -- +(c4);

  \node[draw,circle,inner sep=0pt,minimum size=\radius,name=N1] at (P1) {};
  \draw[thick] (P1) +(c1) -- +(c2) -- +(c3) -- +(c4) -- +(c1) +(c1) -- +(c3) +(c2) -- +(c4);
  \draw[thick] (P1) ++(r1) +(c1) -- +(c2) -- +(c3) -- +(c4) -- cycle;
  \draw[thick,->] (Nempty) -- (N1);
   
  \node[draw,circle,inner sep=0pt,minimum size=\radius,name=N2] at (P2) {};
  \draw[thick] (P2) +(c1) -- +(c2) -- +(c3) -- +(c4) -- +(c1) +(c1) -- +(c3) +(c2) -- +(c4);
  \draw[thick] (P2) ++(r2) +(c1) -- +(c2) -- +(c3) -- +(c4) -- cycle;
  \draw[thick,->] (Nempty) -- (N2);

  \node[draw,circle,inner sep=0pt,minimum size=\radius,name=N3] at (P3) {};
  \draw[thick] (P3) +(c1) -- +(c2) -- +(c3) -- +(c4) -- +(c1) +(c1) -- +(c3) +(c2) -- +(c4);
  \draw[thick] (P3) ++(r3) +(c1) -- +(c2) -- +(c3) -- +(c4) -- cycle;
  \draw[thick,->] (Nempty) -- (N3);
 
  \node[draw,circle,inner sep=0pt,minimum size=\radius,name=N4] at (P4) {};
  \draw[thick] (P4) +(c1) -- +(c2) -- +(c3) -- +(c4) -- +(c1) +(c1) -- +(c3) +(c2) -- +(c4);
  \draw[thick] (P4) ++(r4) +(c1) -- +(c2) -- +(c3) -- +(c4) -- cycle;
  \draw[thick,->] (Nempty) -- (N4);
  
  \node[draw,circle,inner sep=0pt,minimum size=\radius,name=N12] at (P12) {};
  \draw[thick] (P12) +(c1) -- +(c2) -- +(c3) -- +(c4) -- +(c1) +(c1) -- +(c3) +(c2) -- +(c4);
  \draw[thick] (P12) ++(r1) +(c1) -- +(c2) -- +(c3) -- +(c4) -- cycle;
  \draw[thick] (P12) ++(r2) +(c1) -- +(c2) -- +(c3) -- +(c4) -- cycle;
  \draw[thick,->] (N1) -- (N12);
  \draw[thick,->] (N2) -- (N12);
  
  \node[draw,circle,inner sep=0pt,minimum size=\radius,name=N14] at (P14) {};
  \draw[thick] (P14) +(c1) -- +(c2) -- +(c3) -- +(c4) -- +(c1) +(c1) -- +(c3) +(c2) -- +(c4);
  \draw[thick] (P14) ++(r1) +(c1) -- +(c2) -- +(c3) -- +(c4) -- cycle;
  \draw[thick] (P14) ++(r4) +(c1) -- +(c2) -- +(c3) -- +(c4) -- cycle;
  \draw[thick,->] (N1) -- (N14);
  \draw[thick,->] (N4) -- (N14);

\end{tikzpicture}
\end{center}
\caption{\label{fig:dag}DAG visualization of a Markov mesh model in which the sets
$\Lambda=\{\emptyset,\{(0,-1)\},\{(-1,0)\},\{(-1,-1)\},\{(-1,1)\},\{(0,-1),(-1,0)\},\{(0,-1),(-1,1)\}\}$ and 
$\tau=\{(0,-1),(-1,-1),(-1,0),(-1,1)\}$. 
$\boxtimes$ is used in the vertices of the DAG to represent the node $(0,0)$ whilst each $\square$ represents each node $(i,j)\in\lambda$ for
each $\lambda\in\Lambda$.}
\end{figure}

Consider a rectangular lattice of size $m\times n$ and use $v=(i, j), i\in\{1, \dotsc, m\}, j\in\{1,\dotsc,n\}$ to denote a 
specific node of this lattice, corresponding to the notation used for the elements in a matrix. We let $D$ denote the set 
of all nodes, and to each node $v\in D$ we have an associated binary variable $y_v=y_{(i,j)}\in\{0,1\}$. We denote the sequence 
of all these variables by $y = (y_v, v\in D)$, and we use $y_A = (y_v, v\in A)$ to denote the sequence of variables in a 
subset $A\subseteq D$. We define the set of predecessors, $\rho_v=\rho_{(i,j)}$, of a node $v=(i,j)$ to consist of all 
nodes numbered before $(i,j)$ when the nodes are numbered in the lexicographically order, 
i.e. $\rho_v=\rho_{(i,j)}=\{(k,l)\in D : nk + l < ni + j\}$. To each node $v\in D$ the Markov mesh model associates 
a sequential neighborhood $\nu_v\subseteq\rho_v$. Except for nodes close to the boundary of the lattice we 
assume all sequential neighborhoods to be translations of the same template sequential 
neighborhood $\tau\subset\{(i, j) : i, j \in\mathcal{N}, i < 0\}\cup\{(0, j) : j \in\mathcal{N}, j < 0\}$, 
where $\mathcal{N}$ is the set of all integers. A Markov mesh model for $y$ is then assuming the following Markov structure
\begin{equation}\label{eq:mmm}
f(y|\varphi)=\prod_{v\in D}f(y_v|y_{\nu_v},\varphi),
\end{equation}
where $\varphi$ is the model parameters. Moreover, it is assumed that 
\begin{equation}\label{eq:mmm_factor}
f(y_v|y_{\nu_v},\varphi)=\frac{\exp\{y_v\cdot\theta(\delta(\nu_v,y)\ominus v)\}}{1+\exp\{\theta(\delta(\nu_v,y)\ominus v)\}},
\end{equation}
where $\theta(\lambda)$ for $\lambda\subseteq \tau$ is a parameter value associated to the set $\lambda$, 
$\delta(\nu_v,y)=\{v\in\nu_v:y_v=1\}$ is the set of nodes in the sequential neighborhood of node $v$ for which $y_v=1$,
and $\delta(\nu_v,y)\ominus v$ is the set $\delta(\nu_v,y)$ back transformed
to the template sequential neighborhood $\tau$, i.e. for $v=(k,l)$ we have 
$\delta(\nu_v,y)\ominus v=\{(i-k,j-l):(i,j)\in\delta(\nu_v,y)\}$. Letting $\Omega(\tau)$ denote the power set of $\tau$,
\citet{book35} show that $\{\theta(\lambda),\lambda\in\Omega(\tau)\}$ 
can be uniquely represented by a set of interaction parameters $\{\beta(\lambda):\lambda\in\Omega(\tau)\}$ according to the relation
\begin{equation}\label{eq:thetabeta}
\theta(\lambda)=\beta(\lambda)+\sum_{\lambda^\star\subset\lambda}\beta(\lambda^\star).
\end{equation}
To limit the number of free model parameters \citet{tech34} defines a set $\Lambda\subseteq \Omega(\tau)$ of active 
interaction parameters and restrict $\beta(\lambda)=0$ whenever $\lambda\not\in\Lambda$. To facilitate the construction of 
a proposal distribution the set of active interactions $\Lambda$ is restricted to be dense in the sense that if $\lambda\in\Lambda$
one must also have $\lambda^\star\in\Lambda$ for all $\lambda^\star\subset\lambda$. A Markov mesh model is thereby defined
by $\varphi=\{\tau,\Lambda,\{\theta(\lambda):\lambda\in\Lambda\}\}$.  
Note that the set of active interactions $\Lambda$ can be visualized by a DAG, and an example is shown in Figure \ref{fig:dag}.
For a given training image $y$ of interest, \citet{tech34} propose to adopt a Bayesian setting and consider the 
training image as a realization from a Markov mesh model $f(y|\varphi)$ parameterized as discussed above. A 
prior $f(\varphi)$ favoring parsimonious models is constructed, so that the posterior distribution of
interest becomes
\begin{equation}\label{eq:targetdist1}
f(\varphi|y)\propto f(\varphi)f(y|\varphi).
\end{equation}
To sample from this distribution the Metropolis--Hastings algorithm is adopted and two proposal 
distributions tailored to the specific target distribution $f(\varphi|y)$ is constructed.
We use the multiple-try Metropolis--Hastings setup introduced in Section \ref{sec:reversiblejump}
to sample from $f(\varphi|y)$. We adopt the two tailored proposal distributions defined in 
\citet{tech34}. In each iteration we draw at random which of the two proposal strategies to use.
In the next section we briefly describe the proposal distribution and refer to \citet{tech34}
for a more detailed description.

\subsection{\label{subsec:proposal}The tailored proposal distribution}
Two tailored proposal distributions are constructed in \citet{tech34}. One is updating the parameter values
$\{\theta(\lambda):\lambda\in\Lambda\}$ only, whereas the other is proposing changes in all three parts
of $\varphi$. Each time we are to propose a new state $\varphi$ we decide at random what proposal 
distribution to use.

When deciding to update $\{\theta(\lambda):\lambda\in\Lambda\}$ only, 
we keep $\tau$ and $\Lambda$ fixed and generate new parameter values. We first draw a direction 
$\{\Delta(\lambda):\lambda\in\Lambda\}$ from a uniform distribution and let the new parameter values
be defined as $\theta^\star(\lambda)=\theta(\lambda)+\alpha^\star\Delta(\lambda)$, where 
the value of $\alpha^\star$ is sampled from the corresponding full conditional for $\alpha^\star$ in
the target distribution. To generate the sample from the full conditional we use
adaptive rejection sampling as introduced in \citet{pro21}. The resulting proposal 
$\{\theta^\star(\lambda):\lambda\in\Lambda\}$ can be said to be tailored to the specific target 
distribution in question because we sample $\alpha^\star$ from the full conditional.

When choosing to update all three elements of $\varphi$, we propose a change in $\Lambda$ by adding or removing
one element from this set, corresponding to adding or removing one node in the DAG representation illustrated 
in Figure \ref{fig:dag}. We first draw at random whether to add an element to $\Lambda$ or to remove
an element from this set. If it is decided that an element in $\Lambda$ should be removed it is first 
identified what elements in $\Lambda$ that can be removed when requiring also the reduced set to be 
dense. For each of these elements $\lambda^\star\in\Lambda$ we compute the resulting change in the logarithm of
target density by removing $\lambda^\star$ from $\Lambda$ and setting the values of the remaining parameter values
$\{\theta(\lambda):\lambda\in \Lambda\setminus\lambda^\star\}$ by minimizing a sum of squares criterion 
between the current and potential new logarithms of the target densities. As discussed in more detail in \citet{tech34}
the change in the logarithm of the target density when removing $\lambda^\star$ becomes 
$d(\lambda^\star)=\frac{\beta(\lambda^\star)}{2^{|\lambda^{\star}|}}$, where $|\lambda^\star|$ is cardinality of 
$\lambda^\star$. To obtain a tailored proposal we want a higher probability for removing an element
$\lambda^\star$ that results in a small change in target distribution, so we let the 
probability for removing $\lambda^\star$ be 
\begin{equation}\label{eq:proprmv}
q(\lambda^\star)\propto\exp\left\{-\kappa\ \frac{\beta(\lambda^\star)}{2^{|\lambda^\star|}}\right\},
\end{equation}
where $\kappa$ is an algorithmic tuning parameter.

When it is decided to add a new element to $\Lambda$, it must first be decided what $\lambda^\star\not\in\Lambda$ to add. 
As no tailoring is used for this we refer to \citet{tech34} for how this is done. After it has been 
decided that a specific $\lambda^\star$ should be added to $\Lambda$, the associated parameter value $\theta^\star(\lambda^\star)$
must be sampled and potential new values for the old parameters, $\{\theta^\star(\lambda):\lambda\in\Lambda\}$, must be 
decided. As a function of $\theta^\star(\lambda^\star)$ the $\{\theta^\star(\lambda):\lambda\in\Lambda\}$ is chosen by 
adopting the same minimum sum of squares criterion as discussed above. Thereby each $\theta^\star(\lambda),\lambda\in\Lambda$ 
is given 
deterministically as a function of $\theta^\star(\lambda^\star)$. To get a tailored proposal distribution the ideal 
would have been to sample $\theta^\star(\lambda^\star)$ from the full conditional for this value. However,
this full conditional is not available in closed form. It is possible to sample from the full conditional by 
adaptive rejection sampling, but the normalizing constant is not available analytically. Therefore, 
a Gaussian approximation to the full conditional is defined and used as proposal distribution. 
To obtain reasonable values for the mean and variance of this Gaussian proposal distribution, 
a set of samples of $\theta^\star(\lambda^\star)$ is first generated from the full conditional and then 
the sample mean and sample variance is used as mean and variance of the proposal distribution.

\subsection{\label{subsec:experimentalsetup}Experimental setup}
For the target and proposal distributions defined in Sections \ref{subsec:target} and \ref{subsec:proposal}, respectively,
we now want to explore the convergence and mixing properties of the multiple-try reversible jump Metropolis--Hastings algorithm 
defined in Section \ref{sec:reversiblejump}. We run simulation experiments for two graphs. Each of the two graphs are 
characterized by two positive integers $L,N\geq 1$, and given $L$ and $N$ the graph is constructed as follows. 
We start by one node, node $0$ say. We let this node have $N$ neighbors and we say these $N$ neighbors are on level $1$.
To each of the nodes on level $1$ we add $N-1$ more neighbors and say that these $N(N-1)$ nodes are on level $2$. 
Including node $0$ each of the nodes in level $1$ thereby have $N$ neighbors. For the nodes in level $2$, and so on,
we repeat this process until we have defined nodes on level $L$. The resulting graph we denote by $\mathcal{G}_{L,N}$.
The graph in Figure \ref{fig:graph}(a) is a $\mathcal{G}_{2,4}$ graph. In the simulation experiments we use
$\mathcal{G}_{3,5}$ and $\mathcal{G}_{1,1}$ graphs. Note that the $\mathcal{G}_{1,1}$ graph has only two nodes, i.e. only one proposal
in each iteration. The $\mathcal{G}_{3,5}$ has $106$ nodes so that the resulting algorithm uses $105$ proposals in each iteration.

\begin{figure}
        \begin{subfigure}[b]{0.5\textwidth}
                	\vspace*{-0.1cm}
                \includegraphics[width=\linewidth]{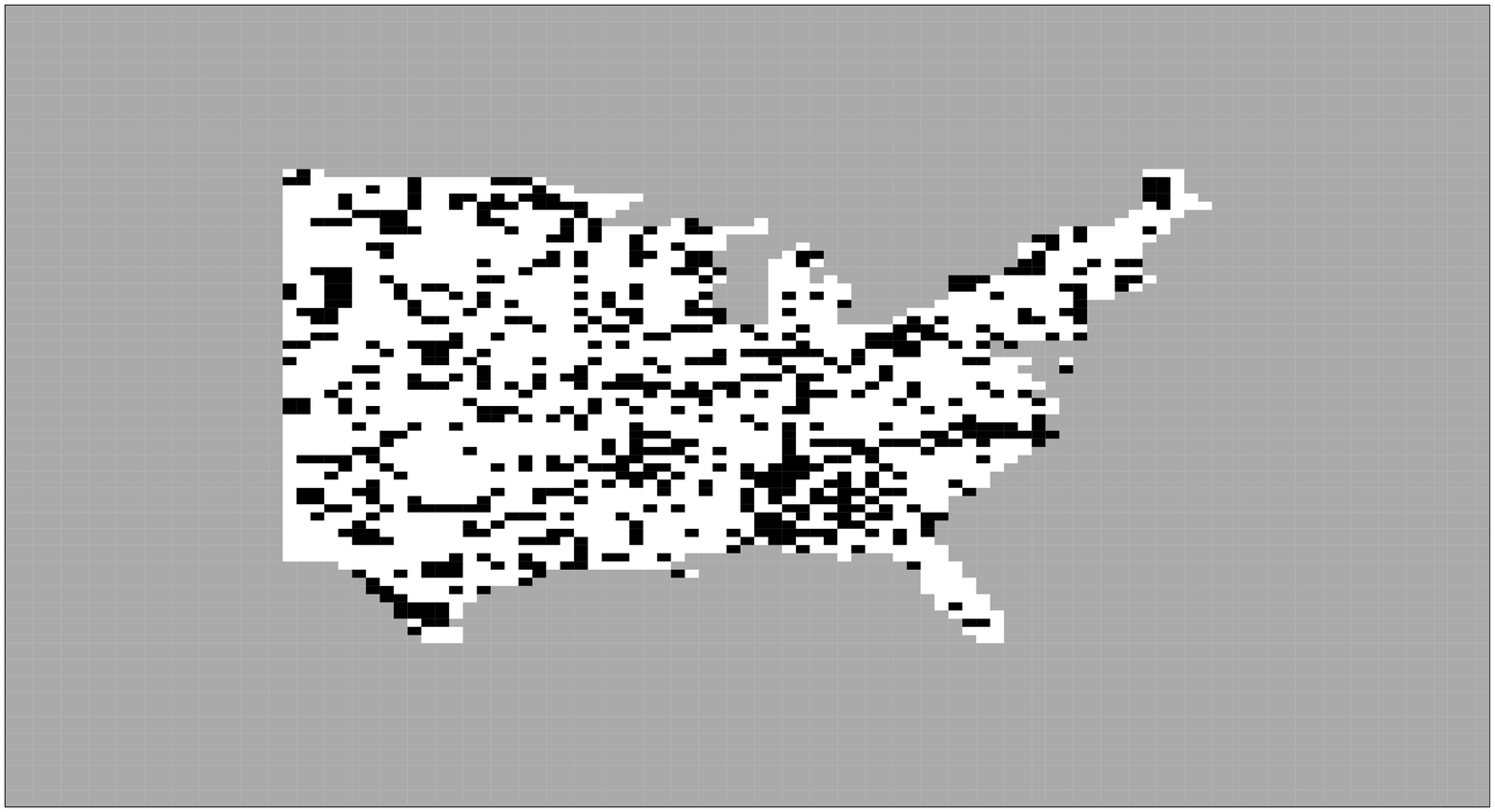}
                	\vspace*{-1.0cm}
                \caption{ }
                \label{fig:cancer}
        \end{subfigure}%
        \hfill
        \begin{subfigure}[b]{0.5\textwidth}
                	\vspace*{-0.1cm}
                \includegraphics[width=\linewidth]{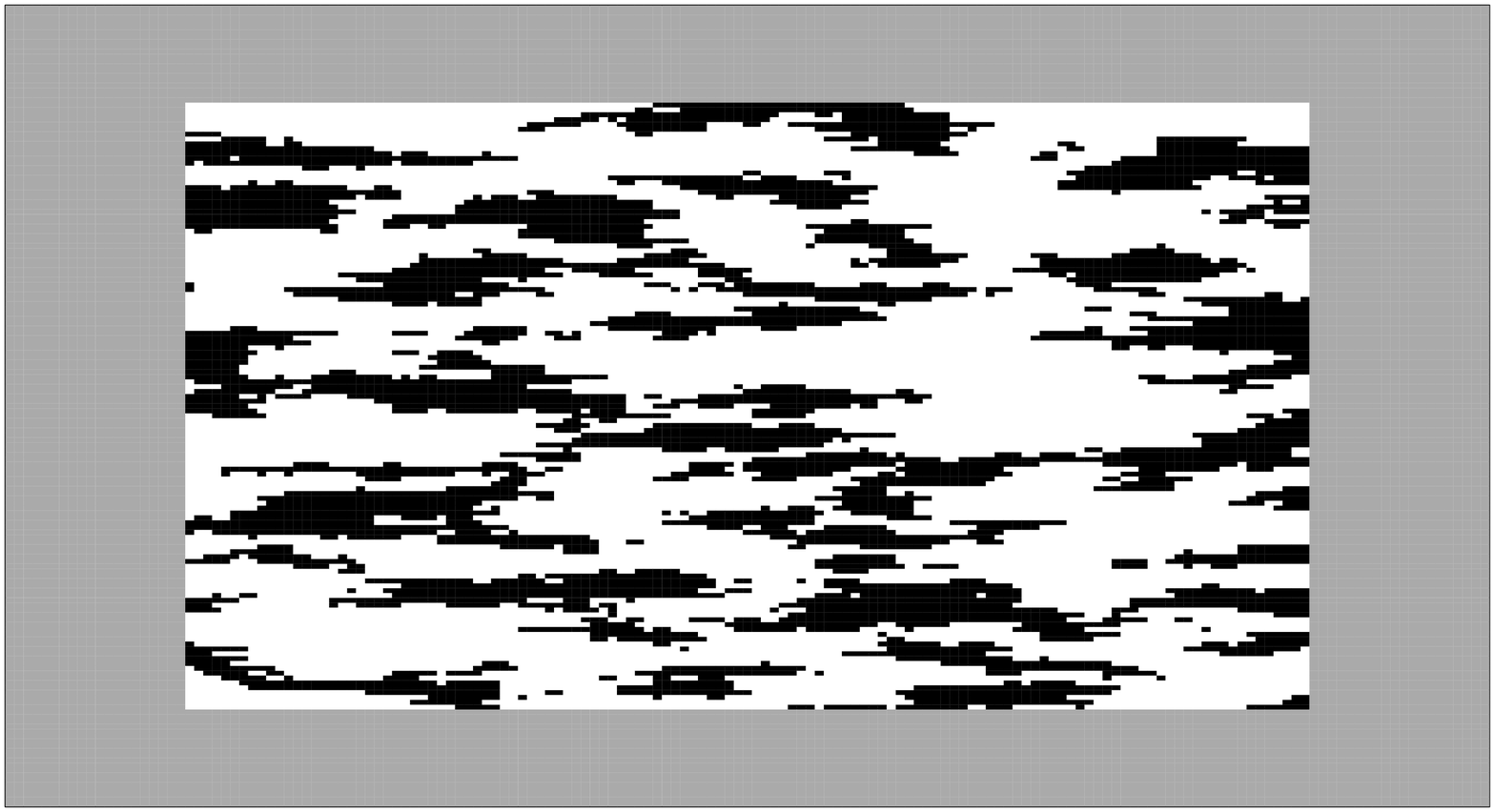}
                	\vspace*{-1.0cm}
                \caption{ }
                \label{fig:sisim}
        \end{subfigure}%
        \caption{Binary training images used in the simulation experiments. The gray area represent unobserved nodes.
          (a) Cancer data set. Black and white pixels 
          represent high and low cancer mortality rates, respectively. (b) Sisim data set.}\label{fig:TI}
\end{figure}

The target distribution we are using is defined for a given training image. We consider two training images, both 
previously considered in \citet{tech34} using a standard Metropolis--Hastings algorithm. 
The first training image, shown in Figure \ref{fig:TI}(a), is a mortality map for liver and gallbladder cancers for 
white males between 1950 and 1959 in the eastern United States, analyzed by \citet{book36}. This data set is
previously considered by \citet{art148}, \citet{art149} and \citet{art158} using Markov random field models, 
see also \citet{book37}. In Figure \ref{fig:TI}(a) the black ($y_v=1$) and white ($y_v=0$) pixels represent counties with high and low 
cancer mortality rates, respectively. Following \citet{tech34}, we define the Markov mesh model on an extended lattice
to reduce the boundary effects. In Figure \ref{fig:TI}(a) this is shown as a gray area which thereby 
represents unobserved nodes.
The second training image we are using is shown in Figure \ref{fig:TI}(b) and is a 
data set previously considered by \cite{art132}. They fitted a Markov mesh model to this data set, but with manually 
chosen neighborhood and interaction structures. This data set was simulated by the sequential indicator
simulation procedure \citep{pro22,book40}, and we name the data set "sisim". The sisim scene is represented on 
a $125\times 125$ lattice. To reduce the border effects
of the Markov mesh model we again include unobserved pixels, shown in gray in Figure \ref{fig:TI}(b).

To simulate from the defined distribution we alternate between the update discussed in Section \ref{sec:reversiblejump} and single 
site Gibbs updates for the values of the unobserved nodes. The parameter space from which we simulate 
is complicated, the dimensionality of the state vector varies and the interpretation of the parameters varies.
To evaluate the convergence and mixing properties of the algorithms we focus on three scalar functions. The two first are 
the number of interactions, i.e. number of elements in the set $\Lambda$, and the logarithm of the 
posterior density. The third scalar function we use is specifically constructed to reveal lack of convergence.
For each of the two graphs $\mathcal{G}_{3,5}$ and $\mathcal{G}_{1,1}$ we make five runs, all starting with the empty model,
$\Lambda=\emptyset$. Separately for each of the two graphs we form the third scalar function as follows.
Based on trace plots of the first two scalar functions we set and discard a (preliminary) burn-in period from each of the 
runs. Based on the simulated models of all five runs we start by finding the most frequently visited model $\Lambda$ and put this 
model into a group number $0$. If the (estimated) probability of this state is less than a threshold $\eta$
we find all visited models $\Lambda$ that can be formed by starting with the model included in group $0$ and thereafter
adding or removing one interaction. We call these models neighbor models of group $0$. The neighbor model with 
the highest (estimated) probability we add to group $0$. If the total frequency of group $0$ is still less than $\eta$, we 
repeat the process. We find all visited neighbor models to models in group $0$, which are not already in group $0$, and 
put into group $0$ the model of these neighbor models with the highest estimated probability. We stop the process 
when the total probability of the models in group $0$ is at least $\eta$ or if the models in the group have no visited  
neighbor models outside the group. We then start form another group of models, group $1$. We first find the most 
frequently visited model which is not in group $0$ and put this model into group $1$. If this model has
probability less than $\eta$ we begin adding visited neighbor models to group $1$ in the same way as described for group $0$, 
except that we now disregard models that are already put into group $0$. Thereafter we make group $2$ in the same way,
now disregarding models that are already in group $0$ or $1$, thereafter we make group $3$ and so on until all 
visited models have been assigned a group. The third scalar function is then defined as the group index of the visited state.
To evaluate whether the chains really have converged we limit the attention to groups that have probabilities larger than or close to 
$\eta$ and find the observed frequencies of the various groups in each of the five runs. If the observed frequencies 
vary a lot it is a clear indication that the chains have not converged.

\subsection{\label{subsec:experimentalresult}Results}
\begin{figure}
        \begin{subfigure}[b]{0.5\textwidth}
         \vspace*{-0.1cm}
                \includegraphics[width=\linewidth]{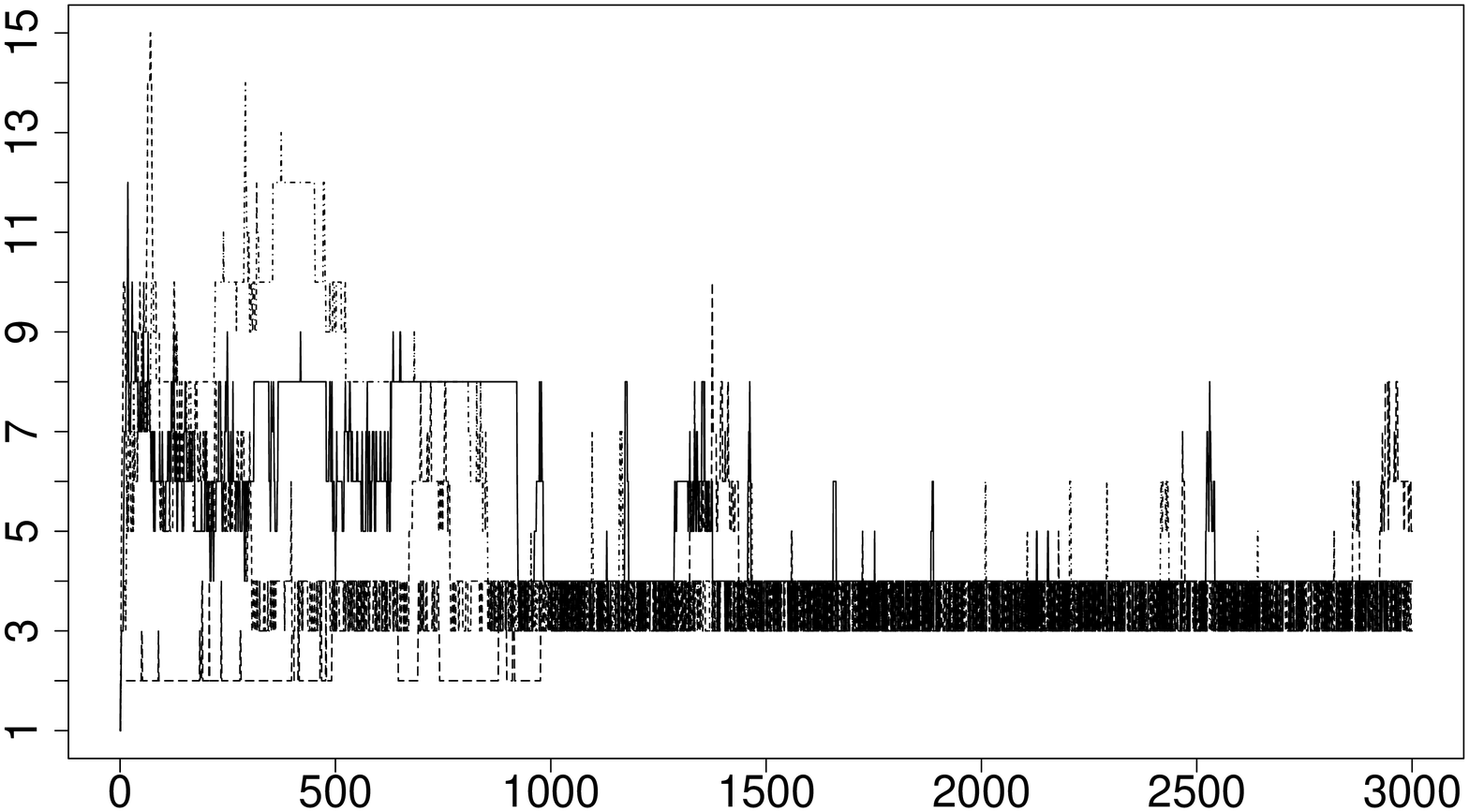}
         \vspace*{-1.0cm}
                \caption[]{}
                \label{fig:multi_num_burnin}
        \end{subfigure}%
        \hfill
        \begin{subfigure}[b]{0.5\textwidth}
        \vspace*{-0.1cm}
                \includegraphics[width=\linewidth]{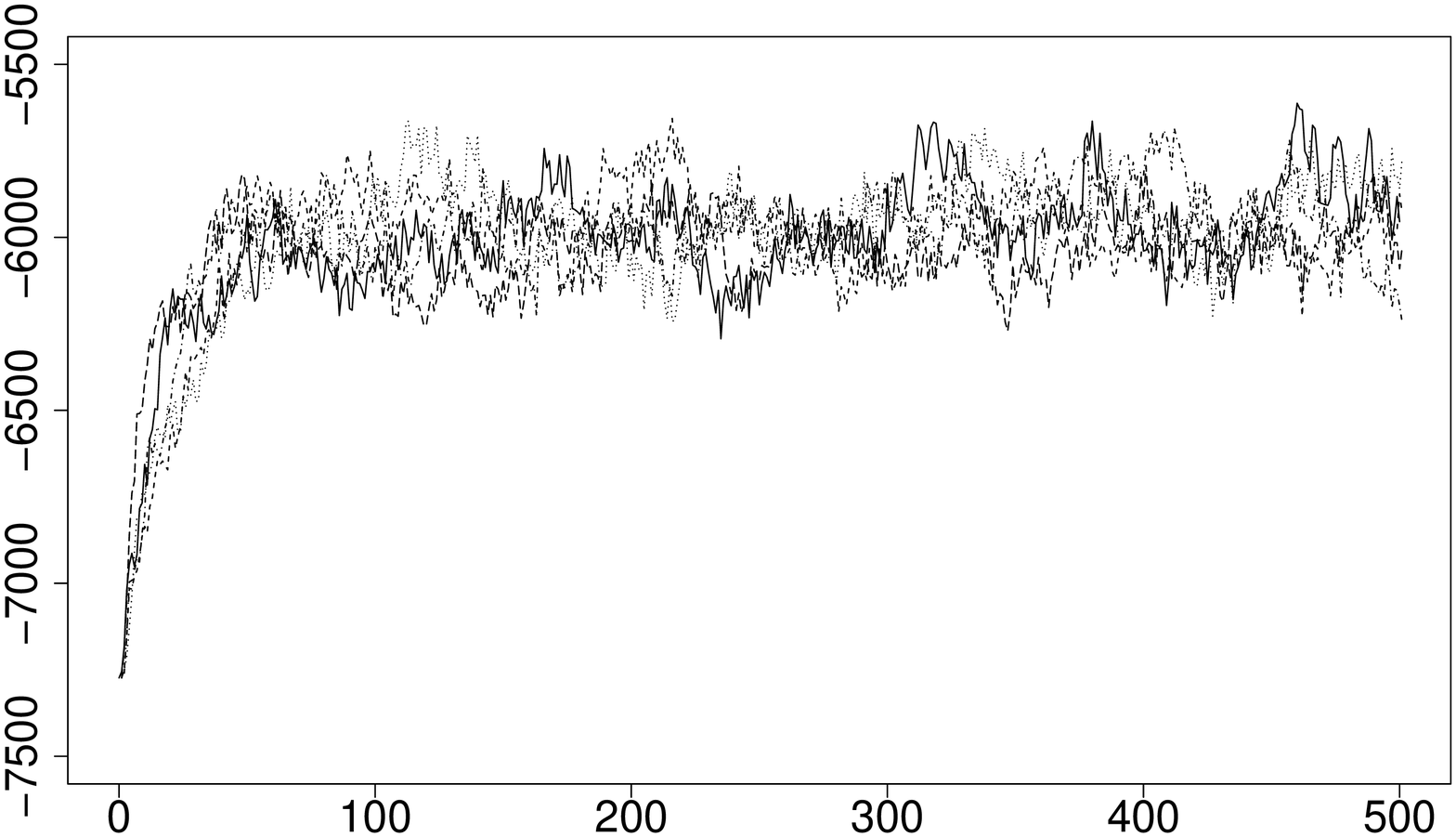}
         \vspace*{-1.0cm}
                \caption[]{}
                \label{fig:multi_post_burnin}
        \end{subfigure}
        
        \begin{subfigure}[b]{0.5\textwidth}
        \vspace*{-0.1cm}
                \includegraphics[width=\linewidth]{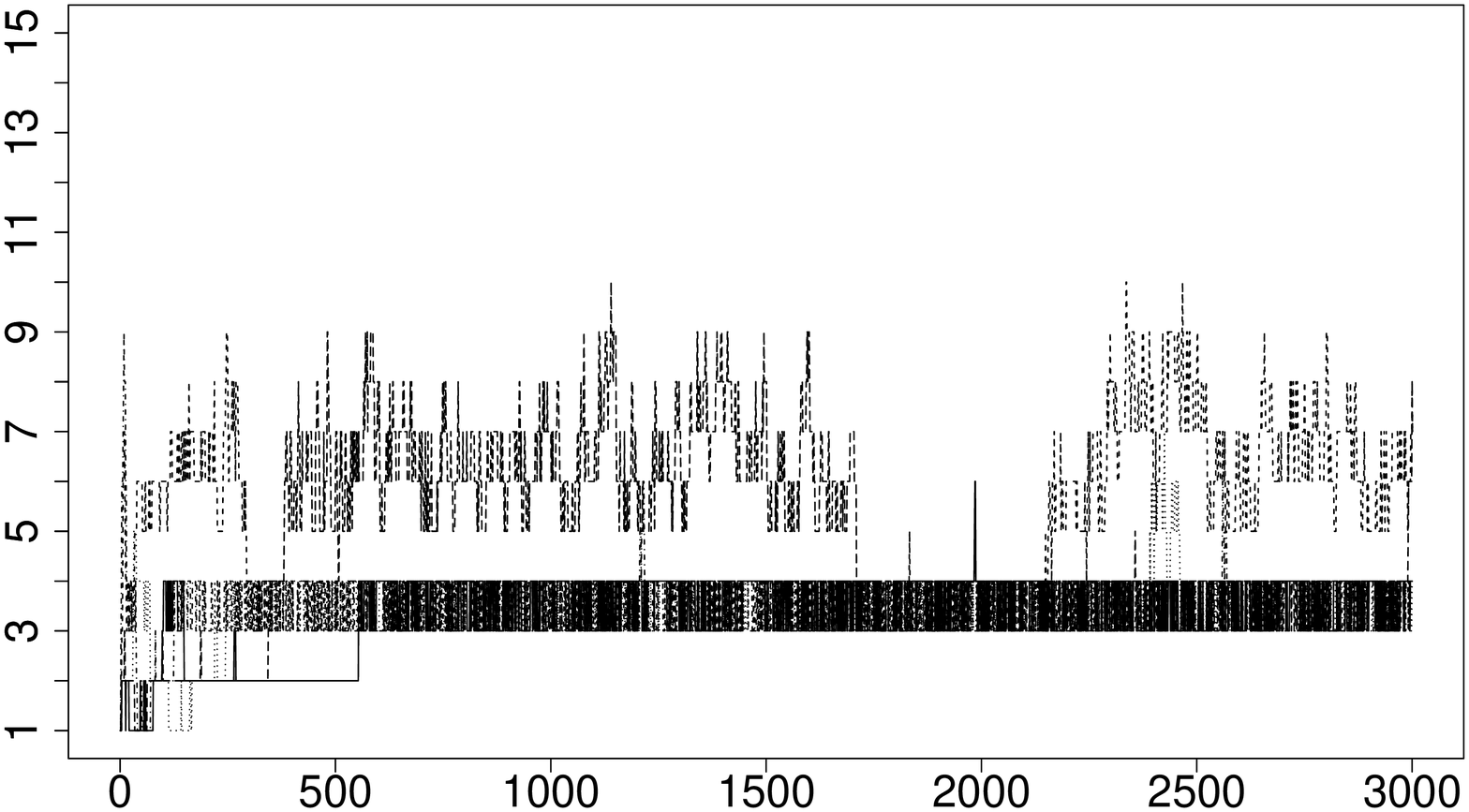}
         \vspace*{-1.0cm}
            \caption[]{}
                \label{fig:two_num_burnin}
        \end{subfigure}%
        \hfill
        \begin{subfigure}[b]{0.5\textwidth}
        \vspace*{-0.1cm}
                \includegraphics[width=\linewidth]{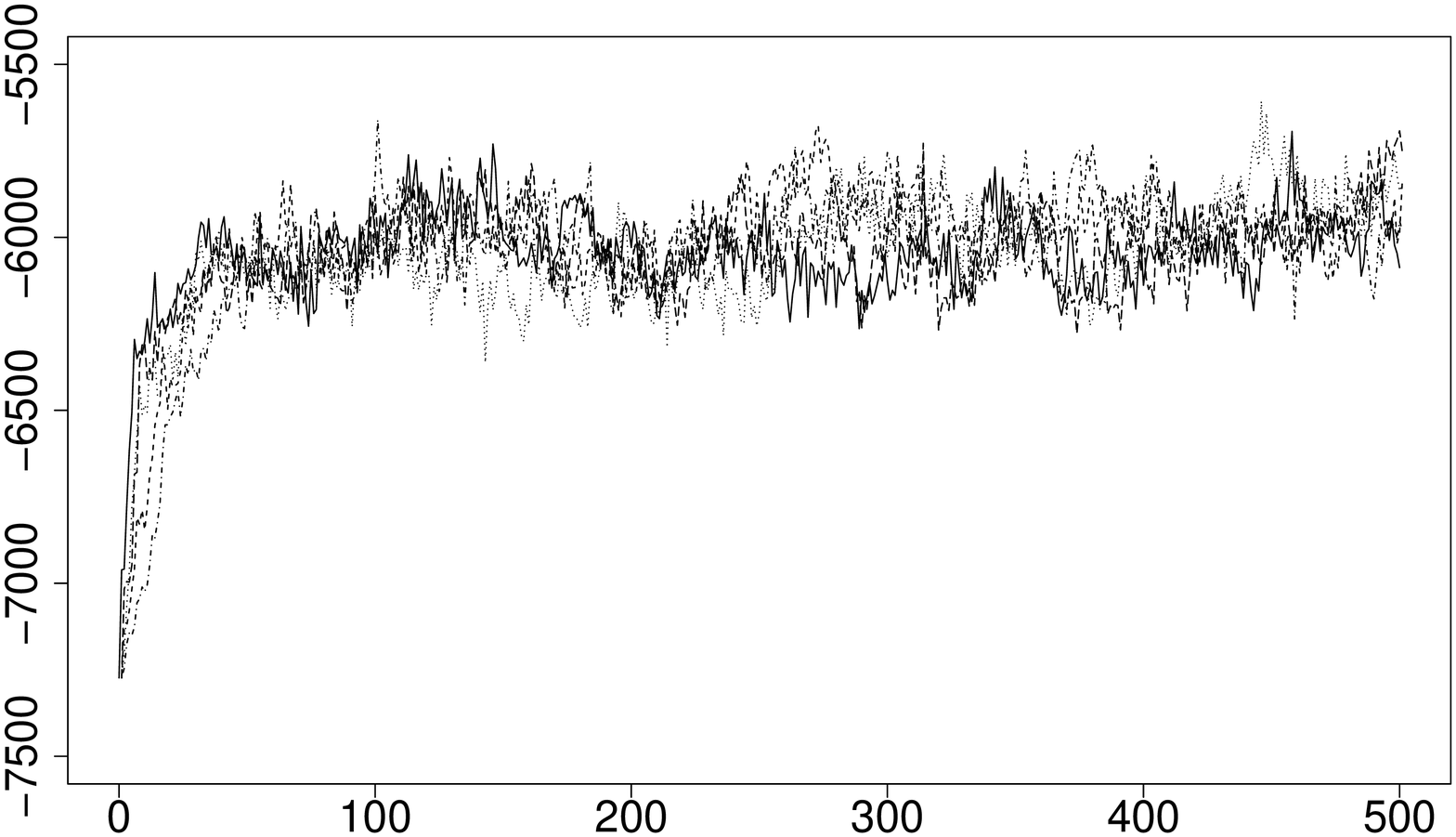}
        \vspace*{-1.0cm}
                \caption[]{}
                \label{fig:two_post_burnin}
        \end{subfigure}
        \caption{Cancer data set example: Trace plots of the first part of the Markov chain runs, where (a) and (b) are simulation results for 
the runs based on the $\mathcal{G}_{3,5}$ graph and (c) and (d) are results for the runs based on the $\mathcal{G}_{1,1}$
graph. The number of interactions is shown in (a) and (c), and the logarithm of the posterior density is shown 
in (b) and (d). In (a) and (b) the number of iterations is specified along the $x$-axis, whereas in 
(c) and (d) the numbers along the $x$-axis is the number of iterations divided by five. All plots show
the traces of five independent runs.}\label{fig:cancer_burnin}
\end{figure}

In this section, we present the simulation results of the setup defined above. We start by showing and discussing the results for 
the cancer data set. We use parallel computing when running based on the $\mathcal{G}_{3,5}$ graph and compare the 
performance of the two algorithms in observed clock time. The run based on the $\mathcal{G}_{1,1}$ graphs is running 
approximately five times 
faster, in clock time, than the run based on the $\mathcal{G}_{3,5}$ graph. Figures \ref{fig:cancer_burnin}(a) and (b)
shows trace plots of the number of interactions and the logarithm of the posterior density for the initial parts of the runs based 
on $\mathcal{G}_{3,5}$. All five runs
are shown in the same plot and the number of iterations is specified along the $x$-axis. The same is shown for the runs based on the 
$\mathcal{G}_{1,1}$ graph in Figures \ref{fig:cancer_burnin}(c) and (d), except that the numbers along the $x$-axis is now the number of 
iterations divided by five so that the results in the two rows are comparable in clock time. Based on these trace plots it is no 
clear difference in the length of the burn-in measured in clock time. Preliminarily we set the length of the burn-in period for the 
$\mathcal{G}_{3,5}$ case to $2000$ iterations and for the $\mathcal{G}_{1,1}$ case to $5\times 2000$ iterations.

We then form groups as discussed in Section \ref{subsec:experimentalsetup}, separately for the $\mathcal{G}_{3,5}$ and $\mathcal{G}_{1,1}$
cases, and estimate the frequencies of each group in each of the five runs. The results for the six most probable groups are shown 
in Table \ref{fig:cancer_group index}.
\begin{table}
  \caption{Cancer data set example: Fractions of the top six most probable group indices for each of five independent runs based on 
    (a) the $\mathcal{G}_{3,5}$ graph, and (b) the $\mathcal{G}_{1,1}$ graph.}\label{fig:cancer_group index}
\begin{center}
\begin{tabular}{ccc}
(a) & ~~~~ &
\begin{tabular}{cccccc}
Group 1: & $0.499$ & $0.481$ & $0.488$ & $0.489$ & $0.463$ \\
Group 2: & $0.392$ & $0.376$ & $0.398$ & $0.390$ & $0.369$ \\
Group 3: & $0.009$ & $0.033$ & $0.011$ & $0.026$ & $0.050$ \\
Group 4: & $0.012$ & $0.035$ & $0.037$ & $0.011$ & $0.033$ \\
Group 5: & $0.030$ & $0.025$ & $0.023$ & $0.025$ & $0.024$ \\
Group 6: & $0.029$ & $0.024$ & $0.016$ & $0.029$ & $0.028$ \\[0.2cm]
\end{tabular}\\
~\\
(b) & ~~~~ & 
\begin{tabular}{cccccc}
Group 1: & $0.505$ & $0.464$ & $0.458$ & $0.479$ & $0.506$ \\
Group 2: & $0.390$ & $0.377$ & $0.370$ & $0.373$ & $0.401$ \\
Group 3: & $0.007$ & $0.047$ & $0.039$ & $0.027$ & $0.006$ \\
Group 4: & $0.022$ & $0.042$ & $0.017$ & $0.037$ & $0.012$ \\
Group 5: & $0.027$ & $0.014$ & $0.040$ & $0.031$ & $0.024$ \\
Group 6: & $0.022$ & $0.021$ & $0.031$ & $0.022$ & $0.029$ \\[0.2cm]
\end{tabular}\\
\end{tabular}
\end{center}
\end{table}
We see that the fractions for the five runs are close to each other, giving a clear indication that the chains for both 
graph cases have converged.

\begin{figure}
        \begin{subfigure}[b]{0.5\textwidth}
         \vspace*{-0.1cm}
                \includegraphics[width=\linewidth]{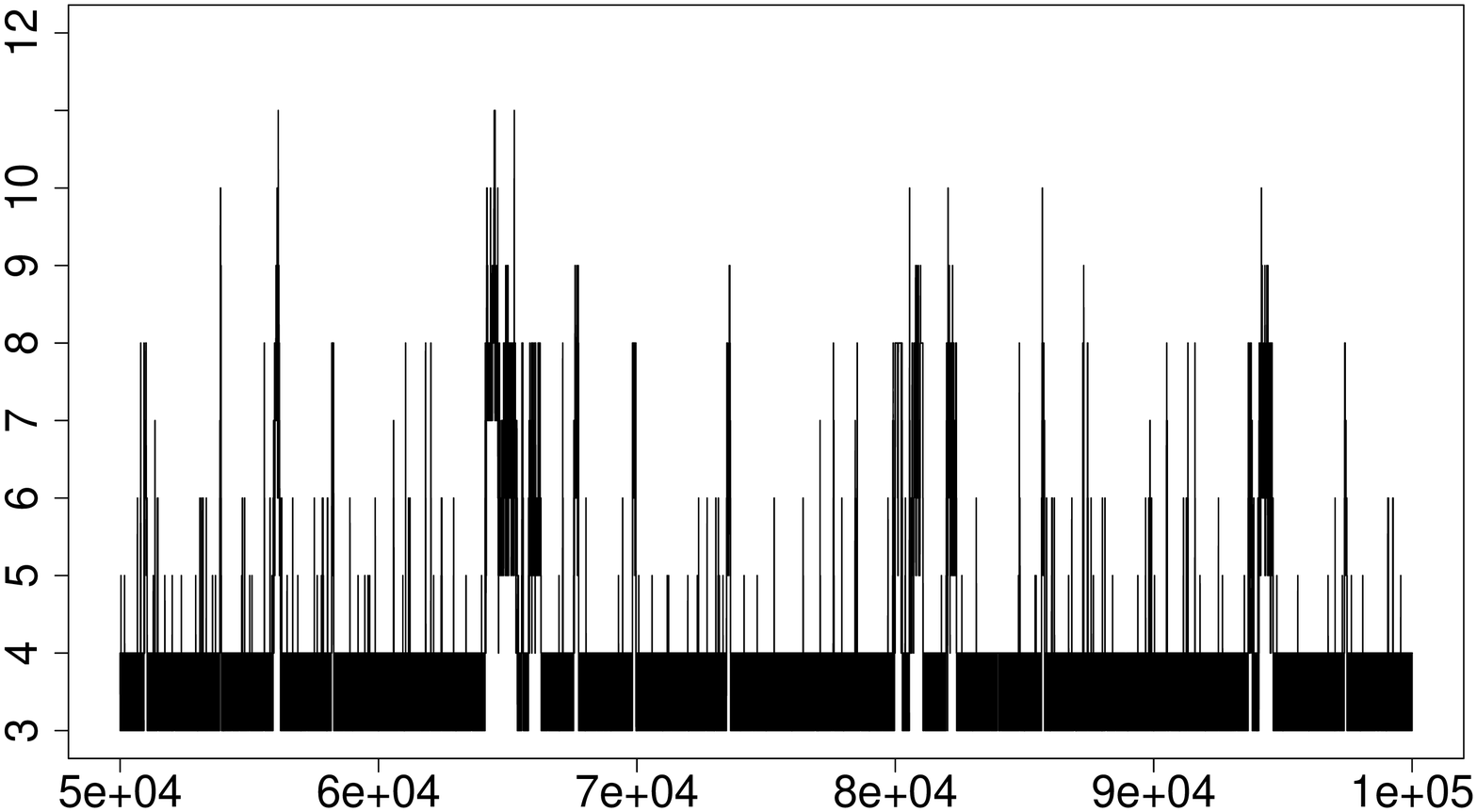}
         \vspace*{-1.0cm}
                \caption[]{}
                \label{fig:multi_num_mixing}
        \end{subfigure}%
        \hfill
        \begin{subfigure}[b]{0.5\textwidth}
        \vspace*{-0.1cm}
                \includegraphics[width=\linewidth]{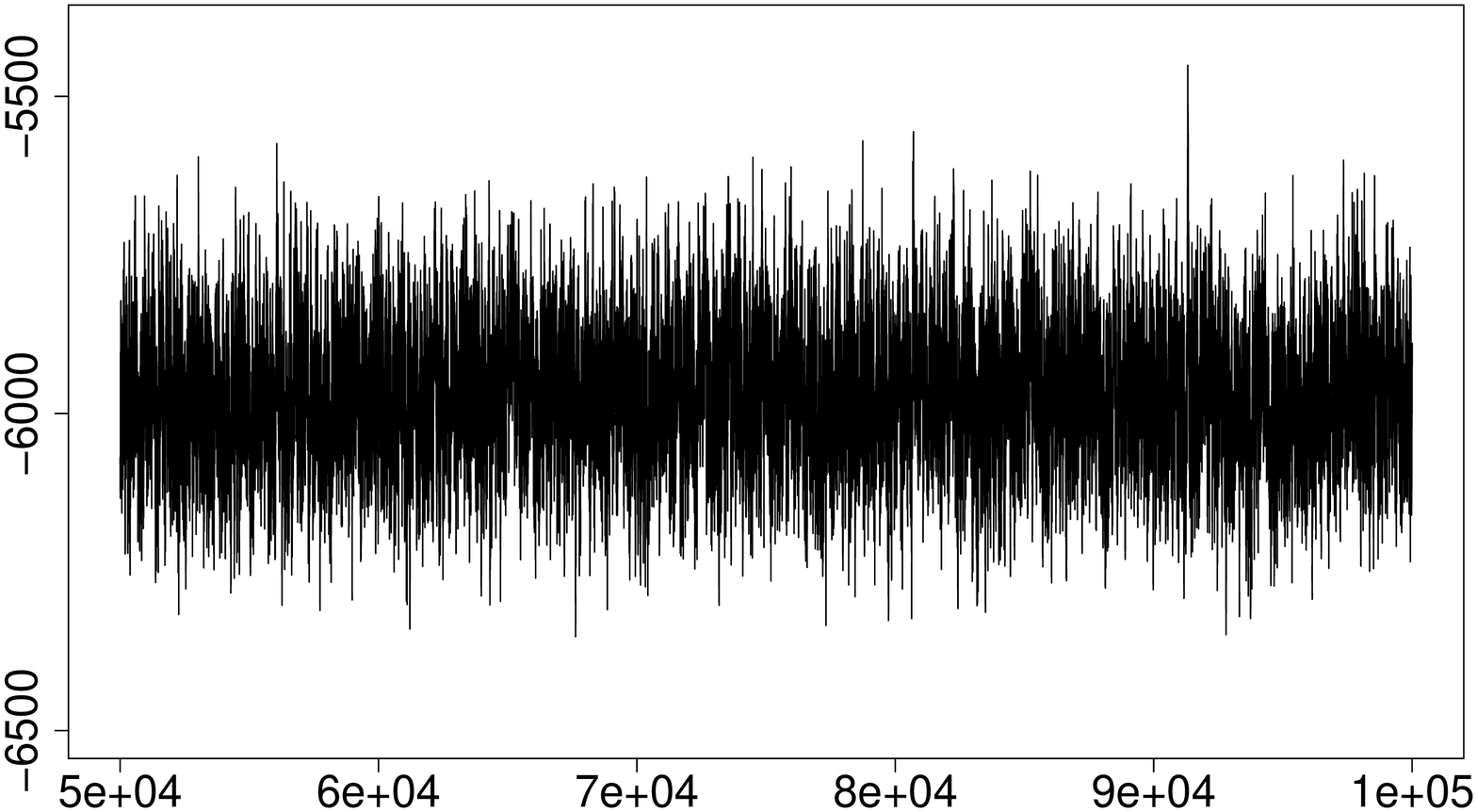}
         \vspace*{-1.0cm}
                \caption[]{}
                \label{fig:multi_post_mixing}
        \end{subfigure}
        
        \begin{subfigure}[b]{0.5\textwidth}
        \vspace*{-0.1cm}
                \includegraphics[width=\linewidth]{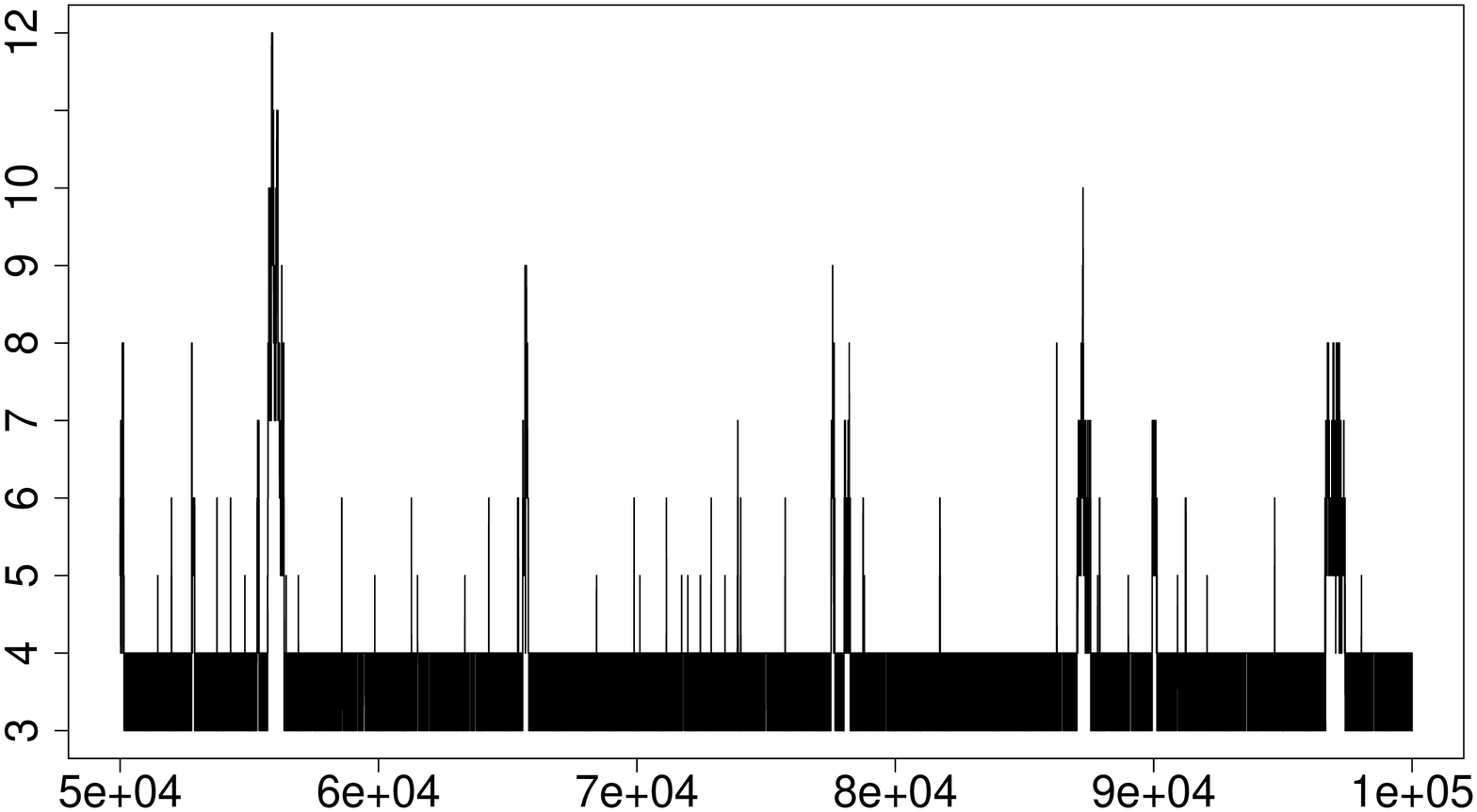}
         \vspace*{-1.0cm}
            \caption[]{}
                \label{fig:two_num_mixing}
        \end{subfigure}%
        \hfill
        \begin{subfigure}[b]{0.5\textwidth}
        \vspace*{-0.1cm}
                \includegraphics[width=\linewidth]{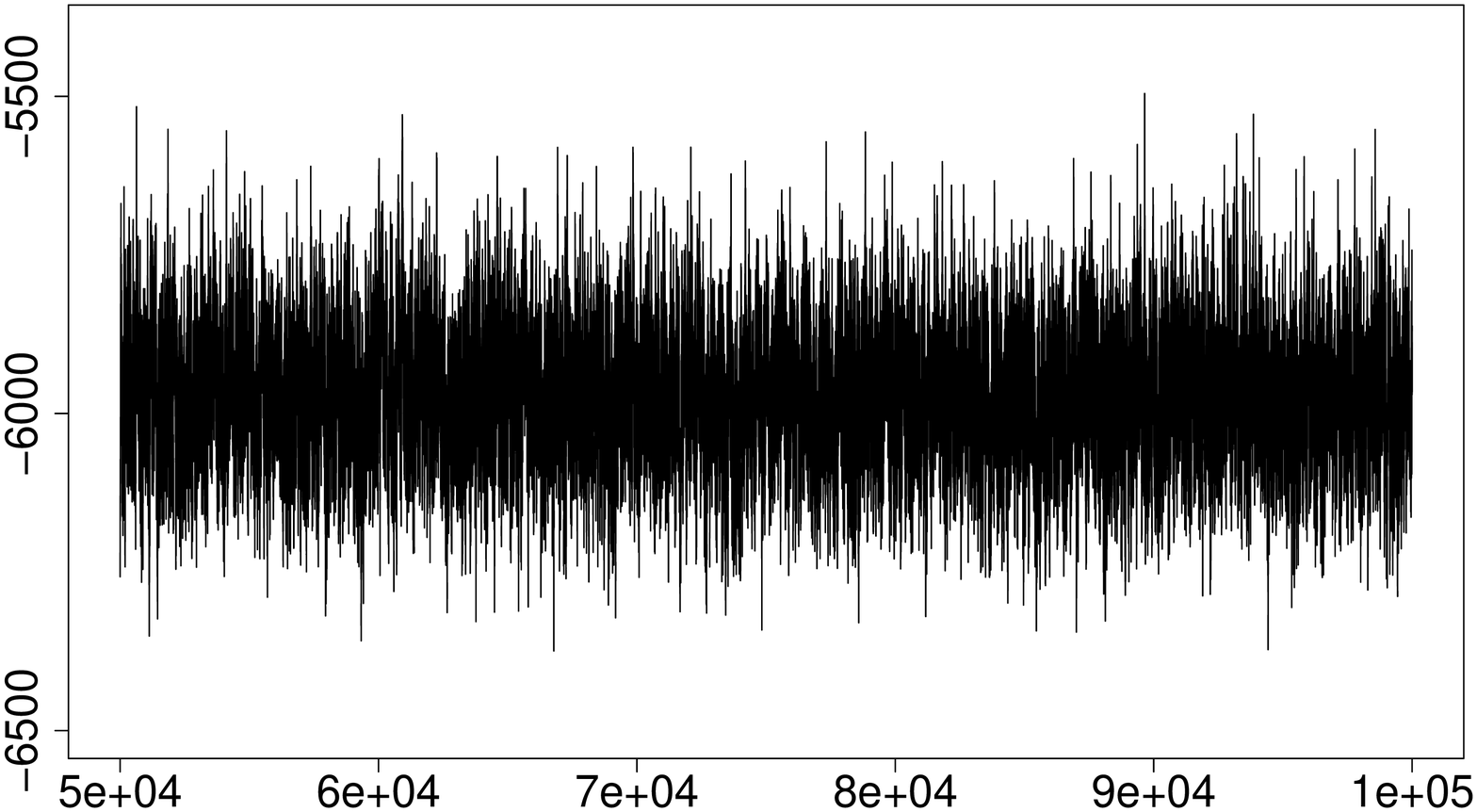}
        \vspace*{-1.0cm}
                \caption[]{}
                \label{fig:two_post_mixing} %
        \end{subfigure}
        \caption{Cancer data set example: Trace plots well after the burn-in period, where (a) and (b) are simulation results for the runs
based on the $\mathcal{G}_{3,5}$ graph and (c) and (d) are results for the runs based on the $\mathcal{G}_{1,1}$ graph.
The number of interactions is shown in (a) and (c), and the logarithm of the posterior density is shown 
in (b) and (d). In (a) and (b) iteration number is specified along the $x$-axis, whereas in 
(c) and (d) the numbers along the $x$-axis is iteration number divided by five. Each trace plot is for 
one Markov chain run.}\label{fig:cancer_mixing}
\end{figure}

\begin{figure}
        \begin{subfigure}[b]{0.5\textwidth}
         \vspace*{-0.1cm}
                \includegraphics[width=\linewidth]{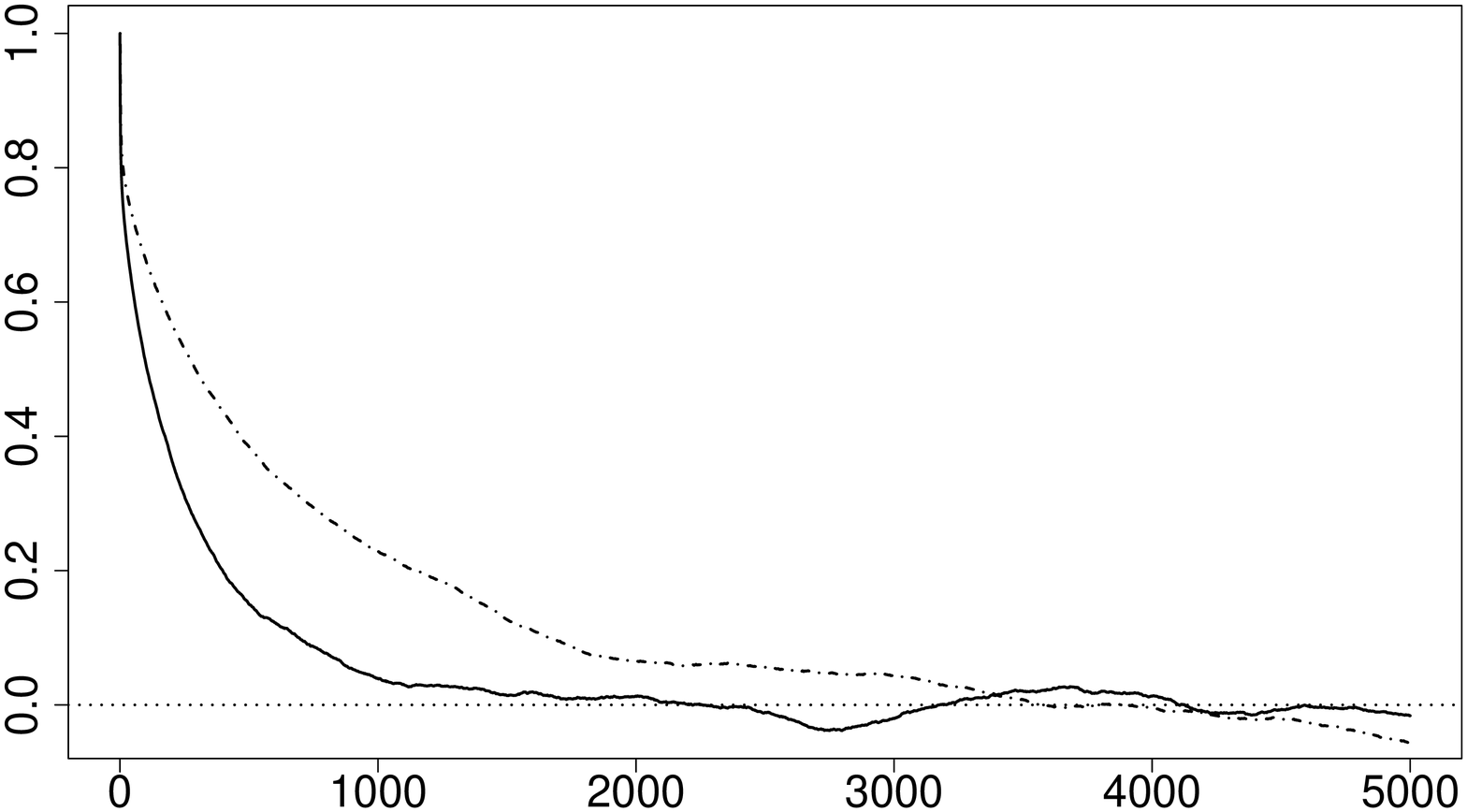}
         \vspace*{-1.0cm}
                \caption[]{}
                \label{fig:num_acf}
        \end{subfigure}%
        \hfill
        \begin{subfigure}[b]{0.5\textwidth}
        \vspace*{-0.1cm}
                \includegraphics[width=\linewidth]{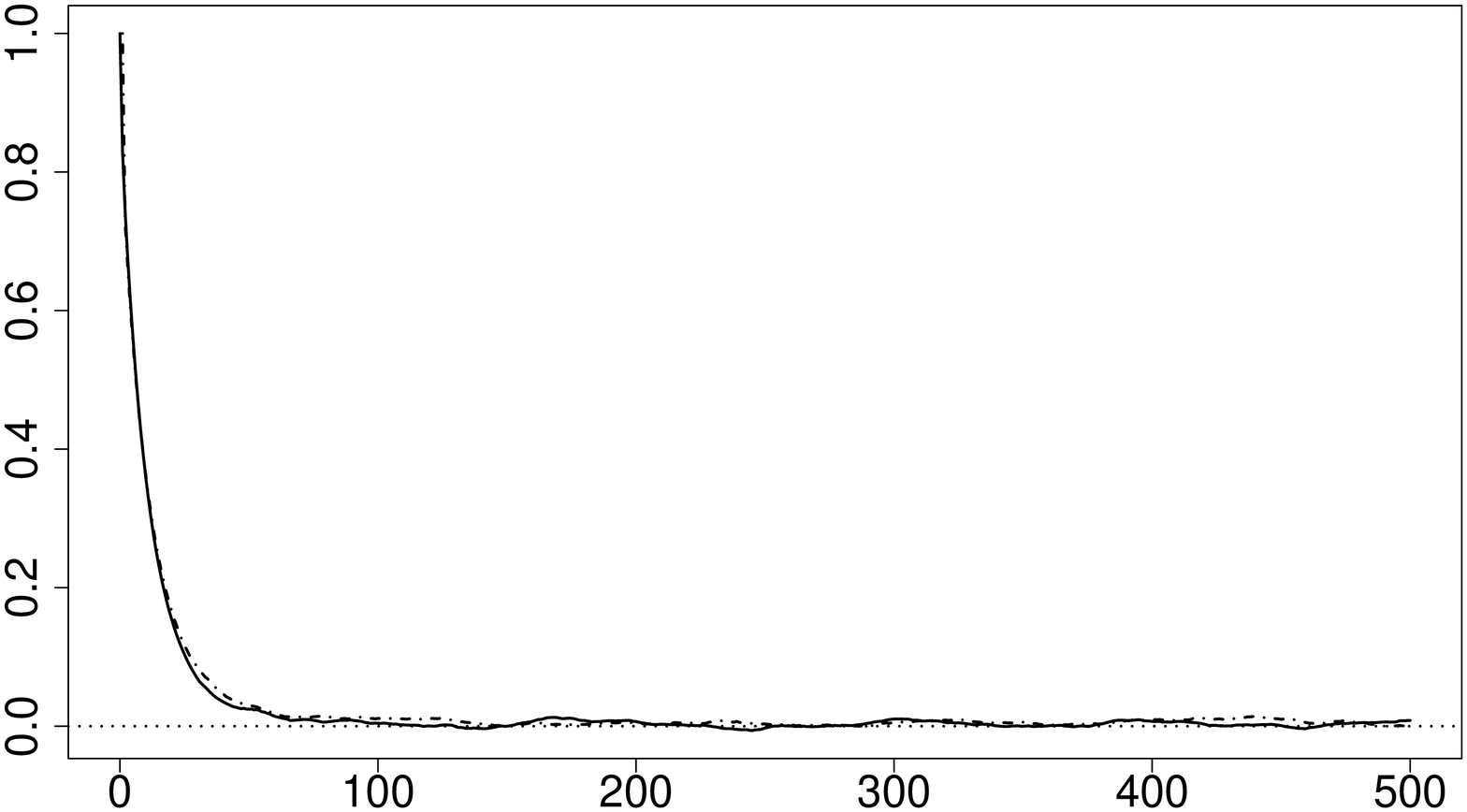}
         \vspace*{-1.0cm}
                \caption[]{}
                \label{fig:post_acf}
        \end{subfigure}
        \caption{Cancer data set example: Estimated autocorrelation functions for (a) number of interactions and (b) logarithm of the posterior 
          density. The solid curves are for the runs based on the $\mathcal{G}_{1,1}$ graph, whereas the dashed 
          curves are for runs based on the $\mathcal{G}_{3,5}$ graph. The numbers along the $x$-axis is 
          the number of iterations for the runs based on the $\mathcal{G}_{3,5}$ graph, whereas for the runs
          based on the $\mathcal{G}_{1,1}$ graph it is the number of iterations divided by five.}\label{fig:cancer_ACF}
\end{figure}

We then shift focus to the mixing properties of the chains. Figure \ref{fig:cancer_mixing}
shows trace plots, well after the burn-in period, of the number of interactions and the logarithm of the 
posterior density for $50000$ iterations for the 
$\mathcal{G}_{3,5}$ case and $5\times 50000$ interactions for the $\mathcal{G}_{1,1}$ case. The number of interactions seem to 
mix better for the $\mathcal{G}_{3,5}$ case than for the $\mathcal{G}_{1,1}$ case, whereas it is difficult to see any difference
for the logarithm of the posterior density. To study the mixing further we also estimate the autocorrelation functions
of the same two scalar functions. The estimates are based on all five runs and shown in Figure \ref{fig:cancer_ACF}. 
The solid curves represent the result for the $\mathcal{G}_{3,5}$ graph case and the dashed curve is the results of the 
$\mathcal{G}_{1,1}$ graph case.
Note that to make the estimated autocorrelation functions comparable (in clock time) the $x$-axis shows the number of 
iterations for the $\mathcal{G}_{3,5}$ case, but the number of iterations divided by five for the $\mathcal{G}_{1,1}$ case.
The results of $\mathcal{G}_{3,5}$ imply clearly better mixing since the corresponding estimated autocorrelation function 
in Figure \ref{fig:cancer_ACF}(a) decays more rapidly.

\begin{figure}
        \begin{subfigure}[b]{0.5\textwidth}
         \vspace*{-0.1cm}
                \includegraphics[width=\linewidth]{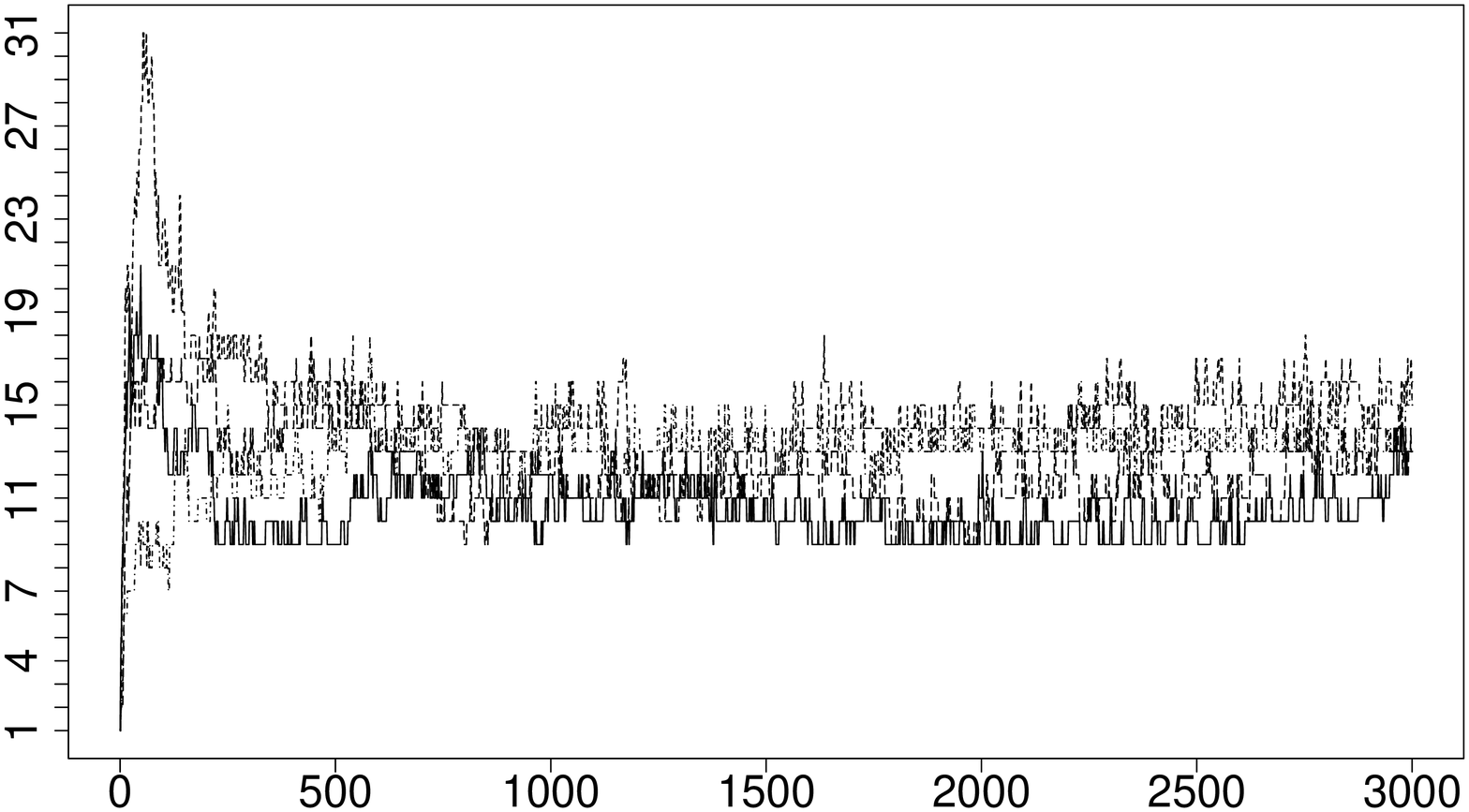}
         \vspace*{-1.0cm}
                \caption[]{}
        \end{subfigure}%
        \hfill
        \begin{subfigure}[b]{0.5\textwidth}
        \vspace*{-0.1cm}
                \includegraphics[width=\linewidth]{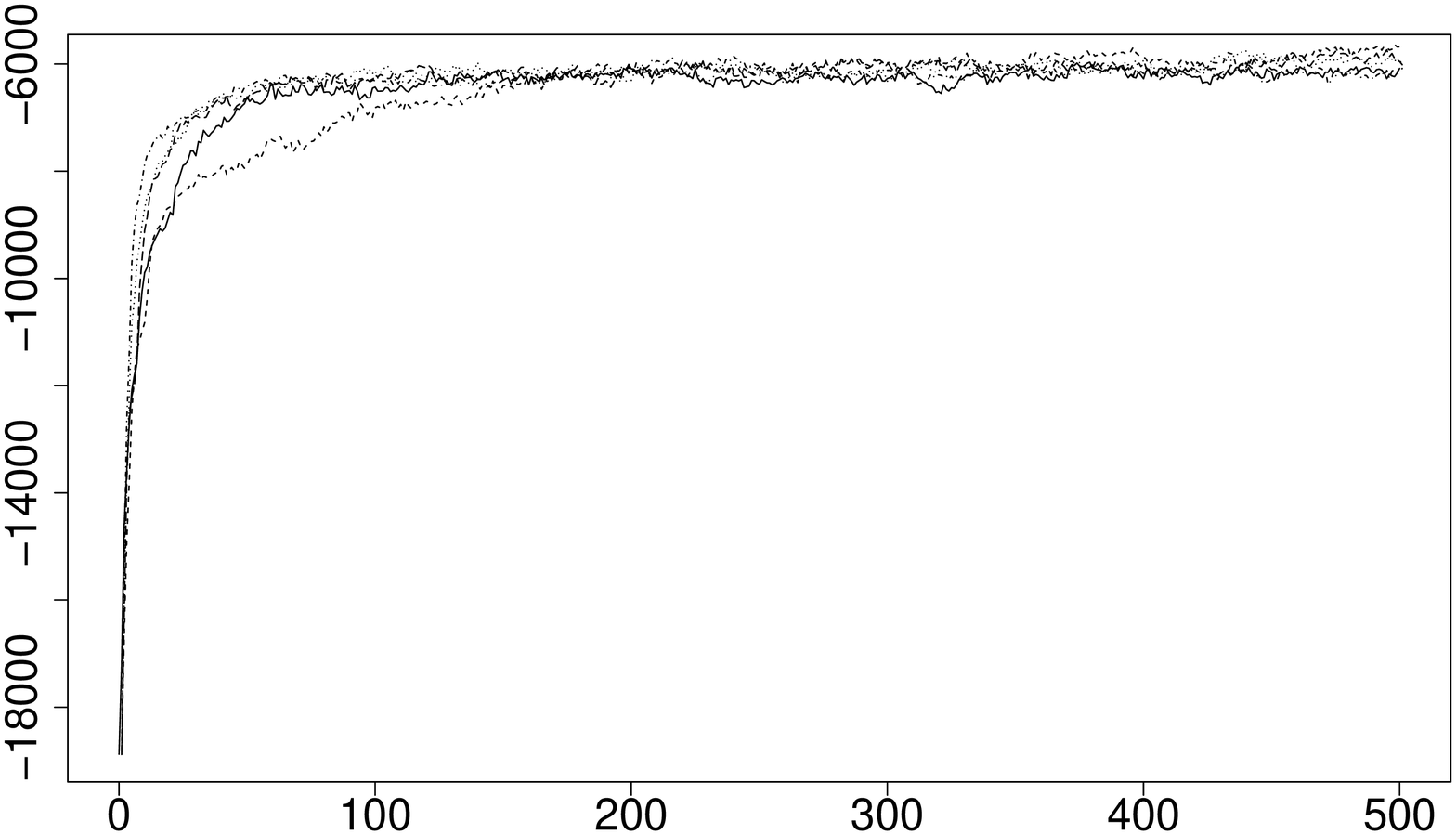}
         \vspace*{-1.0cm}
                \caption[]{}
        \end{subfigure}
        
        \begin{subfigure}[b]{0.5\textwidth}
        \vspace*{-0.1cm}
                \includegraphics[width=\linewidth]{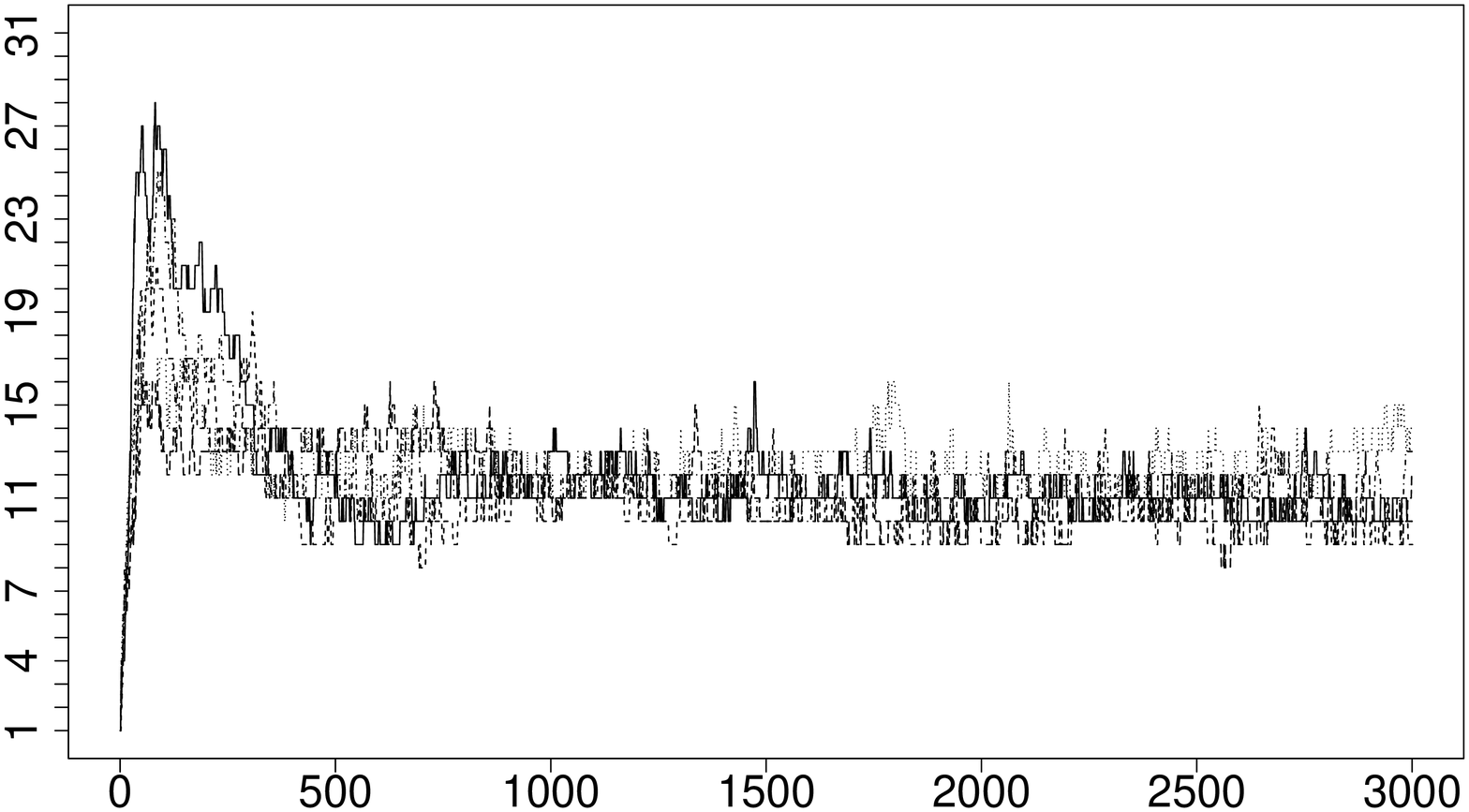}
         \vspace*{-1.0cm}
            \caption[]{}
        \end{subfigure}%
        \hfill
        \begin{subfigure}[b]{0.5\textwidth}
        \vspace*{-0.1cm}
                \includegraphics[width=\linewidth]{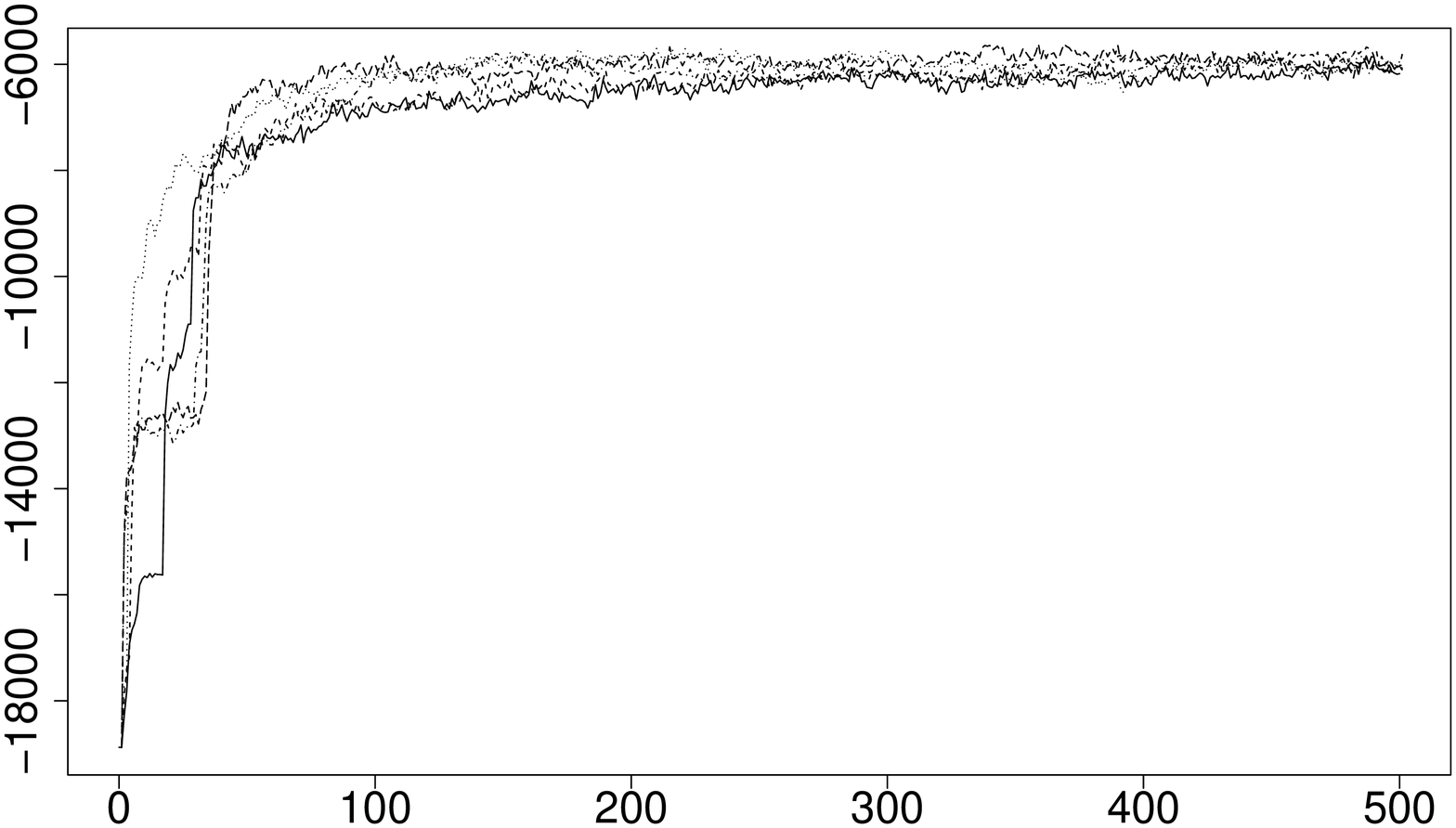}
        \vspace*{-1.0cm}
                \caption[]{}
        \end{subfigure}
\caption{\label{fig:sisim_burnin}Sisim example: Trace plots of the first part of the Markov chain runs, where (a) and (b) 
are simulation results for 
the runs based on the $\mathcal{G}_{3,5}$ graph and (c) and (d) are results for the runs based on the $\mathcal{G}_{1,1}$
graph. The number of interactions is shown in (a) and (c), and the logarithm of the posterior density is shown 
in (b) and (d). In (a) and (b) the number of iterations is specified along the $x$-axis, whereas in 
(c) and (d) the numbers along the $x$-axis is the number of iterations divided by eight. All plots show
the traces of five independent runs.}
\end{figure}

We next present and discuss the results for the sisim training image. With this training image the simulations turned out
to be more troublesome. Again we ran five MCMC runs for each of our two graphs. The runs based on the 
$\mathcal{G}_{1,1}$ graph were approximately a factor eight faster (in clock time) than the runs based on $\mathcal{G}_{3,5}$.
Figure \ref{fig:sisim_burnin} 
corresponds to Figure \ref{fig:cancer_burnin} and the upper row shows trace plots of the number of 
interactions and the logarithm of the posterior density for the initial parts of the runs based on $\mathcal{G}_{3,5}$.
All five runs are shown and the number of interactions is shown along the $x$-axis. Corresponding quantities for 
the runs based on $\mathcal{G}_{1,1}$ are shown in the lower row of Figure \ref{fig:sisim_burnin}, except that the 
numbers along the $x$-axis is here the number of iterations divided by eight. From these trace plots it is difficult to 
see any clear difference in the length of the burn-in periods measured in clock time. Preliminarily we set the length 
of the burn-in period for the $\mathcal{G}_{3,5}$ case to be $2000$ iterations and for the $\mathcal{G}_{1,1}$ case to 
$8\times 2000$ iterations.

\begin{figure}
        \begin{subfigure}[b]{0.5\textwidth}
         \vspace*{-0.1cm}
                \includegraphics[width=\linewidth]{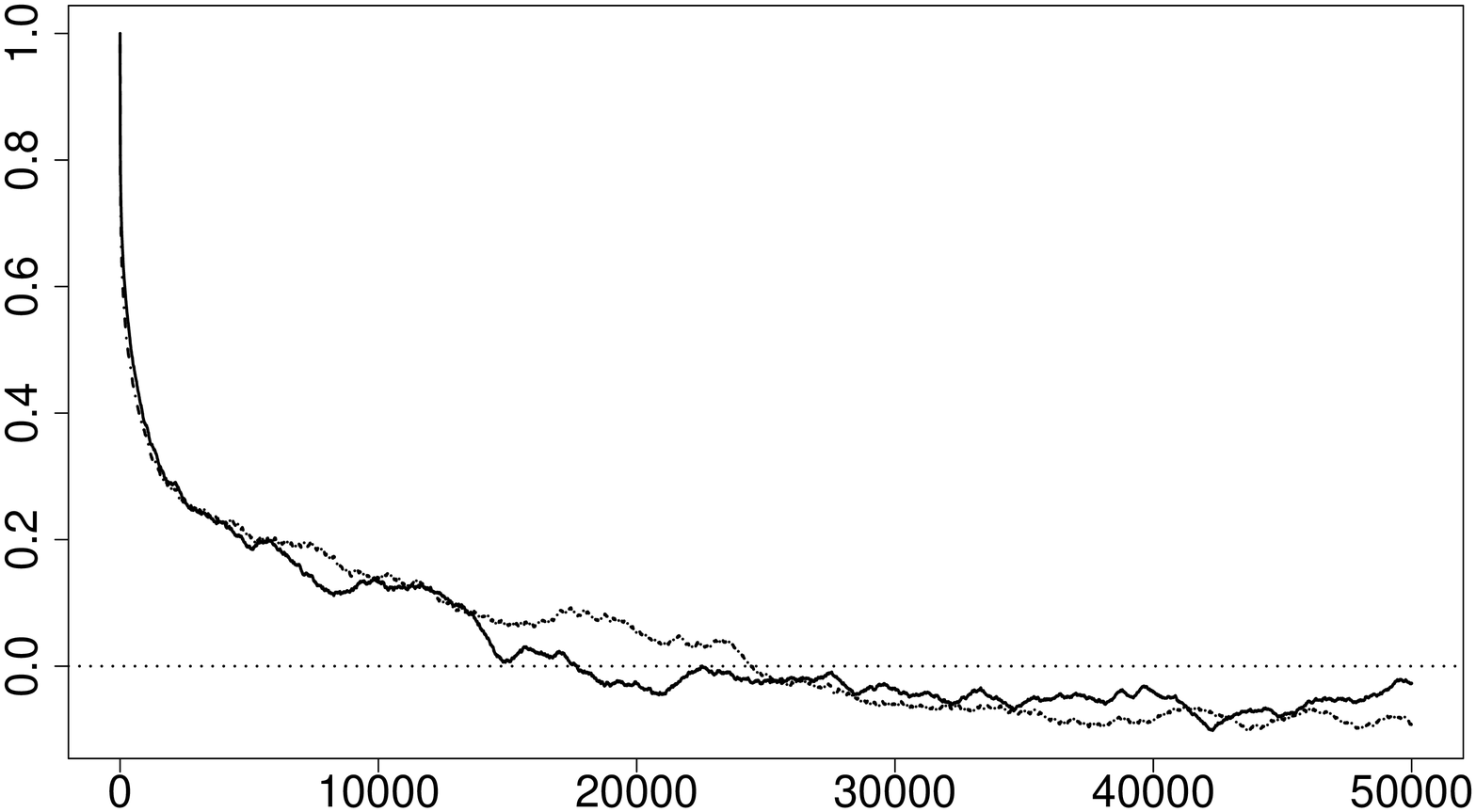}
         \vspace*{-1.0cm}
                \caption[]{}
        \end{subfigure}%
        \hfill
        \begin{subfigure}[b]{0.5\textwidth}
        \vspace*{-0.1cm}
                \includegraphics[width=\linewidth]{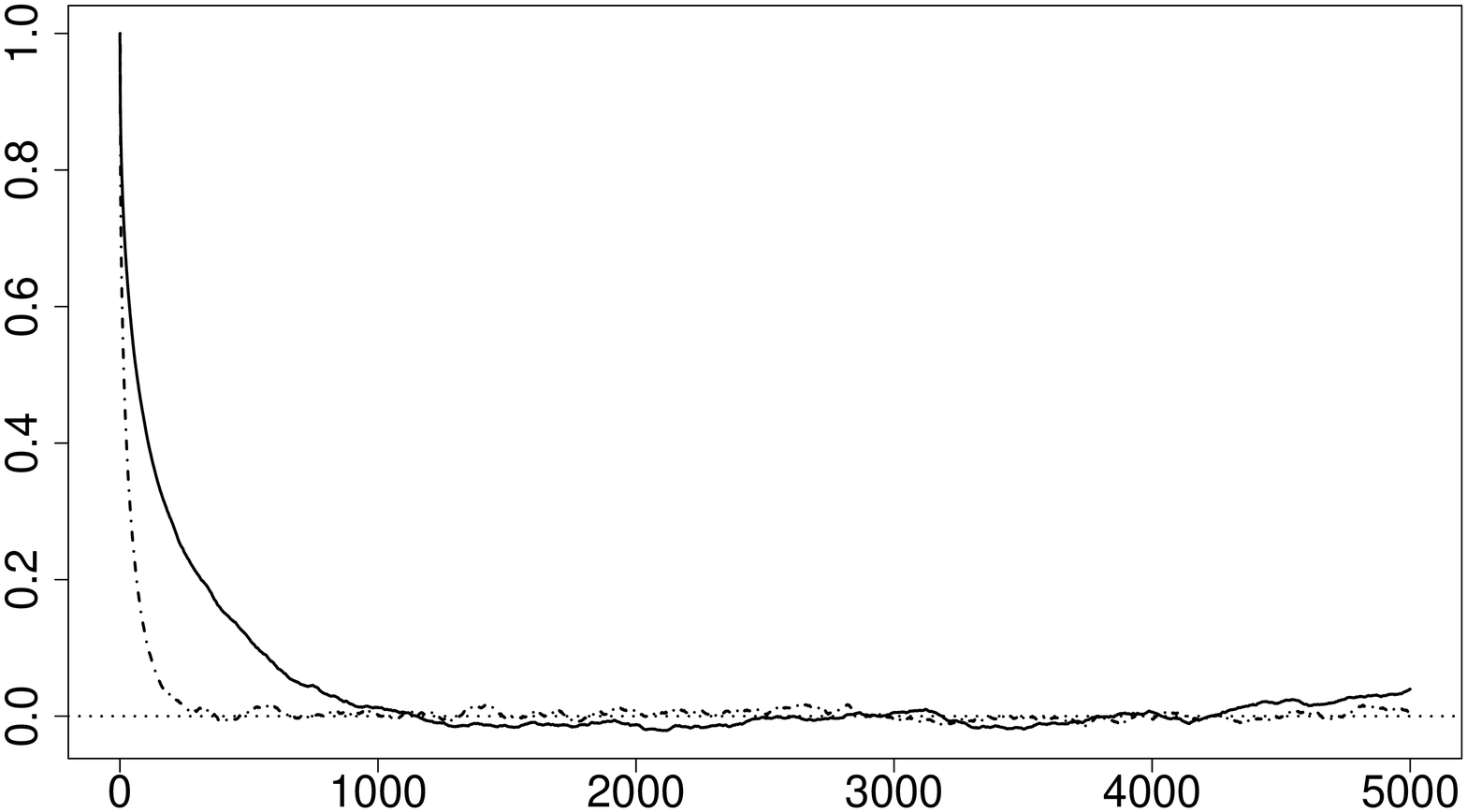}
         \vspace*{-1.0cm}
                \caption[]{}
        \end{subfigure}
\caption{\label{fig:sisim_ACF}Sisim example: Estimated autocorrelation functions for (a) number of 
  interactions and (b) logarithm of the posterior 
  density. The solid curves are for the runs based on the $\mathcal{G}_{1,1}$ graph, whereas the dashed %
  curves are for runs based on the $\mathcal{G}_{3,5}$ graph. The numbers along the $x$-axis is 
  the number of iterations for the runs based on the $\mathcal{G}_{3,5}$ graph, whereas for the runs
  based on the $\mathcal{G}_{1,1}$ graph it is the number of iterations divided by eight.}
\end{figure}

We then again form groups as discussed in Section \ref{subsec:experimentalsetup}, separately for each of the two cases,
and study the frequencies of each group in each of the five runs. The results for the six most probable groups are 
give in Table \ref{tab:sisim_group index}.
\begin{table}
  \caption{\label{tab:sisim_group index}Sisim example: Fractions of the top six most probable group 
    indices for each of five independent runs 
    based on (a) the $\mathcal{G}_{3,5}$ graph, and (b) the $\mathcal{G}_{1,1}$ graph.}
\begin{center}
\begin{tabular}{ccc}
(a) & ~~~ & 
\begin{tabular}{cccccc}
Group 1: & $0.000$ & $0.000$ & $0.121$ & $0.167$ & $0.000$ \\
Group 2: & $0.000$ & $0.109$ & $0.000$ & $0.000$ & $0.150$ \\
Group 3: & $0.000$ & $0.000$ & $0.110$ & $0.189$ & $0.000$ \\
Group 4: & $0.000$ & $0.074$ & $0.000$ & $0.000$ & $0.194$ \\
Group 5: & $0.000$ & $0.208$ & $0.066$ & $0.000$ & $0.000$ \\ %
Group 6: & $0.000$ & $0.000$ & $0.112$ & $0.150$ & $0.000$ \\[0.2cm]
\end{tabular}\\
~\\
(b) & ~~~ & 
\begin{tabular}{cccccc}
Group 1: & $0.116$ & $0.000$ & $0.127$ & $0.033$ & $0.000$ \\
Group 2: & $0.000$ & $0.000$ & $0.005$ & $0.248$ & $0.150$ \\
Group 3: & $0.100$ & $0.000$ & $0.130$ & $0.039$ & $0.000$ \\
Group 4: & $0.000$ & $0.251$ & $0.000$ & $0.000$ & $0.000$ \\
Group 5: & $0.120$ & $0.000$ & $0.103$ & $0.040$ & $0.000$ \\
Group 6: & $0.000$ & $0.000$ & $0.000$ & $0.000$ & $0.258$ \\[0.2cm]
\end{tabular}
\end{tabular}
\end{center}
\end{table}
Both for the runs based on the $\mathcal{G}_{3,5}$ graph and the runs based on $\mathcal{G}_{1,1}$, we 
see that the various runs are not visiting all groups. This clearly indicates very slow mixing and as a 
consequence the preliminarily burn-in periods set is most likely much too short. Since the mixing of the 
simulated Markov chains is so slow it is not possible to get a clear conclusion about the relative
mixing properties of the two Markov chains. However, to get a first indication of the mixing properties for the two chains
we still estimate the autocorrelation functions for the same two scalar functions as used in the cancer data example.
When estimating the autocorrelation functions we discard the burn-in periods preliminarily set as discussed above. 
The estimated autocorrelation function are shown in Figure \ref{fig:sisim_ACF}.
Of the two scalar functions we see that it is the autocorrelations for the number of interactions that decay
slowest and for this scalar function the difference between the two cases is very small. This indicates that the multiple-try 
algorithm based on $\mathcal{G}_{3,5}$ gives approximately the same mixing as for the $\mathcal{G}_{1,1}$ graph 
when using the sisim training image.

\section{\label{sec:closingremarks}Closing remarks}
In this article we define a novel multiple-try Metropolis--Hastings algorithm to be used together with a tailored
proposal distribution. Previously defined multiple-try Metropolis--Hastings algorithms typically generates several 
independent proposals in each iteration, whereas in our setup many of the proposals may be dependent. Moreover,
our multiple-try scheme is new in that the proposals are generated by applying a (tailored) proposal 
mechanism several times without any intermediate acceptance steps. As all multiple-try Metropolis--Hastings
algorithms our setup is also ideal for parallel computing. 

We present two examples to evaluate the effectiveness of our approach. In the examples the target distribution 
is defined on a sample space which is a union of spaces of different dimensions. A reversible jump version of 
our multiple-try algorithm must therefore be used. We adopt a previously defined tailored proposal distribution
and use it in our multiple-try scheme. In one of the examples the multiple-try scheme gives much better mixing properties
compared to a scheme with only one proposal in each iteration, when the two chains are run for the same 
clock time. In the other example the multiple-try scheme does not seem to give any advantages.

A graph $\mathcal{G}$ is used to define our multiple-try Metropolis--Hastings algorithm and the number
of proposals in each iteration is given by the number of nodes in $\mathcal{G}$. We have not yet
explored how the size and form of $\mathcal{G}$ influence the mixing properties of the multiple-try
algorithm. Intuitively we expect it to be beneficial to increase the number of nodes in $\mathcal{G}$
when more processors are available. Moreover, the better tailored the proposal mechanism is 
to the target distribution the more levels $L$ we expect to be optimal.

One should note that our multiple-try scheme can be modified in several ways. We use a Gibbs step to 
generate a new value for $k$, but any proposal distribution may be used for $k$. In particular we
expect it to be advantageous to assign high proposal probabilities to values of $k$ that correspond to 
states that are much different from the current state. Moreover, in the scheme discussed above
we consider the graph $\mathcal{G}$ as given and fixed. By letting also $\mathcal{G}$ be stochastic
one may define a proposal procedure where one on the fly add neighbor nodes to nodes that contain good 
(in some sense) proposals.

\bibliographystyle{jasa}
\bibliography{mybib}

\appendix

\section{\label{app:acceptance}Acceptance probability for the proposal of $k$ in Section \ref{sec:reversiblejump}}
We use notation as in Sections \ref{sec:algorithm} and  \ref{sec:reversiblejump}, 
and let $(k,z)$ and $(\widetilde{k},\widetilde{z})$ denote the current and the proposed states, respectively, 
where $z=(x,\{u_{\{i,j\}}|\{i,j\}\in\mathcal{E}\})$ and $\widetilde{z}=(\widetilde{x},\{\widetilde{u}_{\{i,j\}}|\{i,j\}\in\mathcal{E}\})$. 
Recalling from Section \ref{sec:reversiblejump} that when $k$ and $\widetilde{k}$ are given, there exists a deterministic 
one-to-one relation between $z$ and $\widetilde{z}$. The acceptance probability for the proposal is then
\begin{equation}\label{eq:acceptance_prob}
\alpha(\widetilde{k},\widetilde{z}|k,z)=\min\left\lbrace 1,\dfrac{f(\widetilde{k},\widetilde{x},\{\widetilde{u}_{\{i,j\}}|\{i,j\}\in\mathcal{E}\})r(k)}{f(k,x,\{u_{\{i,j\}}|\{i,j\}\in\mathcal{E}\})r(\widetilde{k})}\cdot \left|\dfrac{\partial\widetilde{z}}{\partial z}\right| \right\rbrace,
\end{equation}
where $\dfrac{\partial\widetilde{z}}{\partial z}$ is the Jacobian determinant for the 
transformation from state $z$ to state $\widetilde{z}$. 

In order to show that the acceptance probability $\alpha(\widetilde{k},\widetilde{z}|k,z)=1$, we need to prove that the value of
\begin{equation}\label{eq:acceptance_ratio}
A(\widetilde{k},\widetilde{z}|k,z) = \dfrac{f(\widetilde{k},\widetilde{x},\{\widetilde{u}_{\{i,j\}}|\{i,j\}\in\mathcal{E}\})r(k)}{f(k,x,\{u_{\{i,j\}}|\{i,j\}\in\mathcal{E}\})r(\widetilde{k})}\cdot \left|\dfrac{\partial\widetilde{z}}{\partial z}\right|
\end{equation}
is identical to $1$. Inserting \eqref{eq:joint_rj} and \eqref{eq:prop_k} into \eqref{eq:acceptance_ratio} and using that $x=x_k$ and
$\widetilde{x}=x_{\widetilde{k}}$, that $u_{\{i,j\}}=u_{(i,j)}$ for $(i,j)\in\mathcal{E}_k$, and 
that $\widetilde{u}_{\{i,j\}}=u_{(i,j)}$ for $(i,j)\in\mathcal{E}_{\widetilde{k}}$, all factors except the Jacobian determinants vanish, so we obtain
\begin{equation}\label{eq:acceptance_ratio1}
\begin{split}
A(\widetilde{k},\widetilde{z}|k,z) &=\dfrac{\prod_{(i,j)\in\mathcal{E}_1\setminus\mathcal{E}_{k}}|J(x_i,u_{(i,j)})|}{\prod_{(i,j)\in\mathcal{E}_1\setminus\mathcal{E}_{\widetilde{k}}}|J(x_i,u_{(i,j)})|}\cdot \left|\dfrac{\partial\widetilde{z}}{\partial z}\right|.
\end{split}
\end{equation}
Trivially, if $k=\widetilde{k}$ we have $A(\widetilde{k},\widetilde{z}|k,z) = 1$. 
In the following we first find $\frac{\partial \widetilde{z}}{\partial z}$ when $k$ and $\widetilde{k}$ are neighbors,
thereafter find the same when $k\neq \widetilde{k}$ and $k$ and $\widetilde{k}$ are not neighbors, and finally we insert these expressions in (\ref{eq:acceptance_ratio1}) to show that $A(\widetilde{k},\widetilde{z}|k,z)$ equals one.

If vertex $k$ and vertex $\widetilde{k}$ are neighbors, then $(k,\widetilde{k})\in\mathcal{E}_{k}$ and $(\widetilde{k},k)\in\mathcal{E}_{\widetilde{k}}$, whereas all other edges in $\mathcal{E}_k$ and $\mathcal{E}_{\widetilde{k}}$ are in the same direction, i.e. $\mathcal{E}_{k}\setminus\{(k,\widetilde{k})\}=\mathcal{E}_{\widetilde{k}}\setminus\{(\widetilde{k},k)\}$, so the 
one-to-one transformation becomes
\begin{equation}\label{eq:onetoone1}
\left. 
    \begin{array}{r}
      \widetilde{x}=g(x,u_{\{k,\widetilde{k}\}}) \\
      \widetilde{u}_{\{k,\widetilde{k}\}}=h(x,u_{\{k,\widetilde{k}\}}) \\[0.1cm]
      \widetilde{u}_{\{i,j\}}=u_{\{i,j\}},(i,j)\in\mathcal{E}_{k}\setminus\{(k,\widetilde{k})\}
    \end{array}\right\}
\mbox{  }  \Leftrightarrow \mbox{  }
  \left\{
    \begin{array}{l}
      x=g(\widetilde{x},\widetilde{u}_{\{k,\widetilde{k}\}}) \\
      u_{\{k,\widetilde{k}\}}=h(\widetilde{x},\widetilde{u}_{\{k,\widetilde{k}\}}) \\[0.1cm]
      u_{\{i,j\}}=\widetilde{u}_{\{i,j\}},(i,j)\in\mathcal{E}_{\widetilde{k}}\setminus\{(\widetilde{k},k)\}.
    \end{array}
  \right.
\end{equation}
Note that the elements of the Jacobian $\dfrac{\partial\widetilde{z}}{\partial z}$ depend on the order of the elements in $z$ and $\widetilde{z}$. 
Without loss of generality, we set $x$ and $u_{\{k,\widetilde{k}\}}$ as the first  and second elements
in $z$, put the remaining variables $u_{\{i,j\}},\{i,j\}\in
\mathcal{E}_k\setminus \{\{ k,\widetilde{k}\}\}$ thereafter in some order,
and arrange the elements in $\widetilde{z}$ correspondingly. The upper
left corner of the Jacobi determinant
$\dfrac{\partial \widetilde{z}}{\partial z}$ then becomes
\begin{equation}
\begin{array}{cc}
\dfrac{\partial g}{\partial x}(x,u_{\{k,\widetilde{k}\}}) &
\dfrac{\partial g}{\partial u}(x,u_{\{k,\widetilde{k}\}}) \\ [0.5cm]
\dfrac{\partial h}{\partial x}(x,u_{\{k,\widetilde{k}\}}) &
\dfrac{\partial h}{\partial u}(x,u_{\{k,\widetilde{k}\}}),
\end{array}
\end{equation}
and the remaining diagonal and non-diagonal elements all become equal to
one and zero, respectively. Thereby we get
\begin{equation}\label{eq:jacobian_neigh}
\dfrac{\partial\widetilde{z}}{\partial z}= J(x,u_{\{k,\widetilde{k}\}}) = J(x_k,u_{(k,\widetilde{k})}),
\end{equation}
where $J(\cdot,\cdot)$ is as defined in \eqref{eq:jacobian} and 
we have used that for the state $(k,z)$ we have $x=x_k$ and $u_{\{k,\widetilde{k}\}}=u_{(k,\widetilde{k})}$.

If $k\neq \widetilde{k}$ and $k$ and $\widetilde{k}$ are not neighbors in the 
graph $\mathcal{G}$, let $k=k_0,k_1,\ldots,k_m=\widetilde{k}$ denote
the shortest path from vertex $k$ to vertex $\widetilde{k}$ in $\mathcal{G}$. For example, if 
$\mathcal{G}$ is the one shown in Figure \ref{fig:graph}(a)
and $k=5$ and $\widetilde{k}=9$, the shortest path has $m=3$, $k_0=5, k_1=1, k_2=3$ and $k_3=9$.
Note that with this notation we also have that $\mathcal{E}_k\setminus\mathcal{E}_{\widetilde{k}} =
\{(k_0,k_1),(k_1,k_2),\ldots,(k_{m-1},k_m)\}$. The transformation from $z$ to $\widetilde{z}$ then may be decomposed into
a series of subtransformations by following the path from $k$ to $\widetilde{k}$ step by step.
Letting $z^s$ denote the state when $k=k_s$, for
$s=0,1,\ldots,m$, we may first transform $z=z^0$ to $z^1$, thereafter transform $z^1$ to $z^2$, and so on until we reach
$z^m=\widetilde{z}$. The Jacobi determinant for the whole series of transformations, 
$\frac{\partial \widetilde{z}}{\partial z}$
is equal to the product of the Jacobi determinants for each of these subtransformations. 
Moreover, since $k_{s-1}$ and $k_s$ by construction are
neighbors in $\mathcal{G}$ for each $s\in\{1,2,\ldots,m\}$ we have from (\ref{eq:jacobian_neigh}) that
\begin{equation}
\frac{\partial z^s}{\partial z^{s-1}} = J(x_{k_{s-1}},u_{(k_{s-1},k_s)}).
\end{equation}
Thereby we get
\begin{equation}
\frac{\partial \widetilde{z}}{\partial z} = \prod_{s=1}^m J(x_{k_{s-1}},u_{(k_{s-1},k_s)}) = 
\prod_{(i,j)\in \mathcal{E}_k\setminus\mathcal{E}_{\widetilde{k}}} J(x_i,u_{(i,j)}).
\end{equation}
Noting that this last expression for $\frac{\partial\widetilde{z}}{\partial z}$ is consistent with 
(\ref{eq:jacobian_neigh}) also when $k$ and $\widetilde{k}$ are neighbors we get for all $k,\widetilde{k}\in\mathcal{V}$,
\begin{equation}\label{eq:A2}
A(\widetilde{k},\widetilde{z}|k,z) = \dfrac{\prod_{(i,j)\in\mathcal{E}_1\setminus\mathcal{E}_{k}}|J(x_i,u_{(i,j)})|}{\prod_{(i,j)\in\mathcal{E}_1\setminus\mathcal{E}_{\widetilde{k}}}|J(x_i,u_{(i,j)})|}\cdot \prod_{(i,j)\in \mathcal{E}_k\setminus\mathcal{E}_{\widetilde{k}}} |J(x_i,u_{(i,j)})|.
\end{equation}
To simplify this expression let $k^\star$ denote the vertex in the shortest path from $k$ to $\widetilde{k}$ that is 
closest to vertex $1$. In particular, $k^\star=1$ if vertex $1$ is in the shortest path from $k$ to $\widetilde{k}$.
We then have $\mathcal{E}_k\setminus\mathcal{E}_{\widetilde{k}} = (\mathcal{E}_k\setminus_{k^\star}) \cup 
(\mathcal{E}_{k^\star}\setminus\mathcal{E}_{\widetilde{k}})$, $\mathcal{E}_1\setminus\mathcal{E}_k = 
(\mathcal{E}_1\setminus\mathcal{E}_{k^\star}) \cup (\mathcal{E}_{k^\star}\setminus\mathcal{E}_k)$ and
$\mathcal{E}_1\setminus\mathcal{E}_{\widetilde{k}} = (\mathcal{E}_1\setminus\mathcal{E}_{k^\star}) \cup
(\mathcal{E}_{k^\star}\setminus\mathcal{E}_{\widetilde{k}})$, which can be used to split in two each of the three products in 
(\ref{eq:A2}). We then get
\begin{eqnarray}
A(\widetilde{k},\widetilde{z}|k,z) &=& \frac{
\prod_{(i,j)\in\mathcal{E}_1\setminus\mathcal{E}_{k^\star}} |J(x_i,u_{(i,j)})|}{
\prod_{(i,j)\in\mathcal{E}_1\setminus\mathcal{E}_{k^\star}} |J(x_i,u_{(i,j)})|}\nonumber
\times \frac{\label{eq:A3}
\prod_{(i,j)\in\mathcal{E}_{k^\star}\setminus\mathcal{E}_{k}} |J(x_i,u_{(i,j)})|}{
\prod_{(i,j)\in\mathcal{E}_{k^\star}\setminus\mathcal{E}_{\widetilde{k}}} |J(x_i,u_{(i,j)})|}\\
&\times& \left[\prod_{(i,j)\in\mathcal{E}_k\setminus\mathcal{E}_{k^\star}} |J(x_i,u_{(i,j)})|\right] 
\times \left[\prod_{(i,j)\in\mathcal{E}_{k^\star}\setminus\mathcal{E}_{\widetilde{k}}} |J(x_i,u_{(i,j)})|\right] \\
&=& \frac{\prod_{(i,j)\in\mathcal{E}_{k^\star}\setminus\mathcal{E}_{k}} |J(x_i,u_{(i,j)})|}{
\prod_{(i,j)\in\mathcal{E}_k\setminus\mathcal{E}_{k^\star}} |J(x_i,u_{(i,j)})|}\nonumber
\end{eqnarray}
Moreover, first using (\ref{eq:jacobianInverse}) and thereafter that $(i,j)\in\mathcal{E}_k\setminus\mathcal{E}_{k^\star}
\Leftrightarrow (j,i)\in\mathcal{E}_{k^\star}\setminus\mathcal{E}_k$ we get
\begin{equation}
\frac{1}{\prod_{(i,j)\in\mathcal{E}_k\setminus\mathcal{E}_{k^\star}} |J(x_i,u_{(i,j)})|} = 
\prod_{(i,j)\in \mathcal{E}_k\setminus\mathcal{E}_{k^\star}} |J(x_j,u_{(j,i)})| = 
\prod_{(i,j)\in\mathcal{E}_{k^\star}\setminus\mathcal{E}_k} |J(x_i,u_{(i,j)}|.
\end{equation}
Inserting this in (\ref{eq:A3}) we see that all factors cancel and we get $A(\widetilde{k},\widetilde{z}|k,z)=1$.
The proof is thereby complete.

\end{document}